\documentclass{aa}

\usepackage{graphicx}
\usepackage{txfonts}
\usepackage[]{hyperref} 
\usepackage{booktabs} 
\usepackage{array} 
\usepackage{paralist} 
\usepackage{verbatim} 
\usepackage{float}
\restylefloat{table}
\usepackage{tabularx}
\usepackage{xspace}
\usepackage[caption = false]{subfig}
\usepackage{longtable}
\usepackage{textcomp}
\usepackage{siunitx}
\usepackage{lscape}
\usepackage{pdflscape}
\usepackage{afterpage}
\usepackage{comment}

\DeclareSIUnit \AA {\text{\r{A}}}

\newcommand{\pysme}{PySME\xspace}
\newcommand{\idlsme}{IDL~SME\xspace}
\newcommand{\epseri}{$\epsilon$~Eri\xspace}
\newcommand{\cnc}{55~Cnc\xspace}

\newcommand{\hnpeg}{HN~Peg\xspace}
\newcommand{\hdten}{HD~102195\xspace}
\newcommand{\hdthirteen}{HD~130322\xspace}
\newcommand{\hdseventeen}{HD~179949\xspace}
\newcommand{\hdeighteen}{HD~189733\xspace}
\newcommand{\wasp}{WASP~18\xspace}

\newcommand{\uteff}{\si{\kelvin}}
\newcommand{\ulogg}{\ensuremath{\log(\si{\centi\meter\per\second\squared})}}
\newcommand{\umonh}{\si{dex}}
\newcommand{\uvmic}{\si{\kilo\meter\per\second}}
\newcommand{\uvmac}{\si{\kilo\meter\per\second}}
\newcommand{\uvsini}{\si{\kilo\meter\per\second}}

\newcommand{\steff}{\ensuremath{T_\text{eff}}\xspace}
\newcommand{\slogg}{\ensuremath{\log g}\xspace}
\newcommand{\smonh}{[M/H]\xspace}
\newcommand{\svmic}{\ensuremath{v_\text{mic}}\xspace}
\newcommand{\svmac}{\ensuremath{v_\text{mac}}\xspace}
\newcommand{\svsini}{\ensuremath{v~\sin i}\xspace}
\newcommand{\srv}{\ensuremath{v_\mathrm{rv}}\xspace}

\NewDocumentCommand {\val} { m o o }{%
\IfValueT{#2}{%
\IfValueT{#3}{\ensuremath{#1^{+#2}_{-#3}}}%
\IfNoValueT{#3}{\num{#1 \pm #2}}}%
\IfNoValueT{#2}{\num{#1}}%
}

\newcommand{\opt}[1]{\texttt{#1}}

\usepackage{natbib}
\bibpunct{(}{)}{;}{a}{}{,} 
\graphicspath{{./img/}}

\makeatletter
\renewcommand*\aa@pageof{, page \thepage{} of \pageref*{LastPage}}
\makeatother

\addto\extrasenglish{%
    
}

\addto\extrasenglish{%
    
}

\begin{document} 

   \title{\pysme}

   \subtitle{Spectroscopy Made Easier}

   \author{A. Wehrhahn\inst{1} \and N. Piskunov\inst{1} \and T. Ryabchikova\inst{2}}

   \institute{Department of Physics and Astronomy, Uppsala University, Box 516, 75120 Uppsala, Sweden
   \and Institute of Astronomy of the Russian Academy of Sciences, Pyatnitskaya str. 48, 119017, Moscow, Russia}

   \date{Received \today; accepted ?}

 
  \abstract
   {The characterization of exoplanets requires reliable determination of the fundamental parameters of their host stars. Spectral fitting plays
   an important role in this process. For the majority of stellar parameters, matching synthetic spectra to the observations provides a robust
   and unique solution for fundamental parameters, such as effective temperature, surface gravity, abundances, radial and rotational
   velocities and others.}
   {Here we present a new software package for fitting high resolution stellar spectra that is easy to use, available for common platforms
   and free from commercial licenses. We call it \pysme. It is based on the proven Spectroscopy Made Easy (later referred to as \idlsme or
   "original SME") package.}
   {The IDL part of the original SME code has been rewritten in Python, but we kept the efficient C++ and FORTRAN code responsible for
   molecular-ionization equilibrium, opacities and spectral synthesis. In the process we have updated some components of the optimization
   procedure offering more flexibility and better analysis of the convergence. The result is a more modern package with the same functionality
   of the original SME.}
   {We apply \pysme to a few stars of different spectral types and compared the derived fundamental parameters with the results from \idlsme
   and other techniques. We show that \pysme works at least as well as the original SME.}
   {}

   \keywords{techniques: spectroscopic -- methods: data analysis -- methods: numerical -- stars: fundamental parameters -- stars: solar-type} 

   \maketitle
%

\section {Introduction}
%
%

The determination of the fundamental stellar properties is important in many different fields of astronomy, from galactic archaeology to exoplanet studies. High resolution spectroscopy is one of the most reliable ways to determine these parameters, as it analyses the absorption features of the spectrum based on first principles. Especially with ever increasing sizes of surveys it becomes necessary to have analysis tools that can be applied to a large number of targets as well. Here we present the most recent evolution in the popular Spectroscopy Made Easy package (SME, \citealt{1996A&AS..118..595V,2017A&A...597A..16P}) -- \pysme. Previous iterations of SME were limited by the closed source IDL language in its application to large samples, this has been solved here by transitioning to Python. Additionally this allowed us to improve on some other aspects of the software as well.

\pysme is one of only a few stellar spectral analysis tools available, others include Turbospectrum \citep{1993ApJ...418..812P,2012ascl.soft05004P,2022arXiv220600967G}, MOOG \citep{1973ApJ...184..839S,2012ascl.soft02009S}, Korg \citep{2022arXiv221100029W}, SYNSPEC \citep{2011ascl.soft09022H,2017arXiv170601859H,2021arXiv210402829H}, SYNTHE \citep{1993sssp.book.....K,2004MSAIS...5...93S}, and SPECTRUM \citep{1994AJ....107..742G}.

In this paper we focus our analysis on a relatively small sample of target stars to illustrate the changes, assess the performance, and discuss some improved practical aspects of installation, use and support of this new incarnation of SME. The target stars selected here are all exoplanet host stars as those are of particular interest to exoplanet studies, but of course this new version of SME can be applied to large variety of stars, just as the existing SME.

It should be noted that exoplanet host stars have statistically higher metallicities than their non exoplanet hosting counterparts \citep{2003AJ....126.2015H,2005ApJ...622.1102F}.

The paper is divided into four parts. We start by introducing \pysme, highlighting the changes and the differences from the original IDL implementation (\autoref{sec:pysme}). Then we perform a comparison between the new \pysme and the \idlsme implementation (\autoref{sec:comparison}), to test both the accuracy of the spectral synthesis as well as the stability of the stellar parameter fitting. Afterwards we present a spectroscopic analysis carried out for a sample of planet-hosting FGK stars (\autoref{sec:analysis}) and compare the results to those of other studies (\autoref{sec:comparison_results}). Finally we analyse trends in the derived parameters and their uncertainties in relation to the stellar effective temperature (\autoref{sec:trends}), as a proxy for the stellar type.

\section{\pysme}
\label{sec:pysme}
SME is a spectral synthesis and fitting tool for interpretation of stellar spectra. While we focus here on exoplanet host stars exclusively, \pysme is applicable to a wide variety of stars. In synthesis mode it generates
a spectrum for a given set of stellar parameters (\steff, \slogg, abundances, macro- and micro-turbulence \svmac
and \svmic, and projected equatorial rotation velocity \svsini) for specified spectral intervals and spectral
resolution (instrumental profile). In analysis mode SME finds the optimal values for selected stellar parameters
that result in the best fit to the provided observations given the uncertainties of the data. The free
parameters may include one or several (or even all) stellar parameters from the list above, including
specific elemental abundances as well as some atomic line parameters of any transition in the line list.

The original SME package was written in IDL with the radiative transfer solver inherited from the antique FORTRAN
code \opt{Synth} by Piskunov \citep{1992pess.conf...92P} accessed in the form of an external dynamic library.
Throughout the years more and more improved and updated parts of spectral synthesis were incorporated into the
library, which was later named the SME library\footnote{\url{https://github.com/AWehrhahn/SMElib}}.
It currently includes the molecular-ionisation equilibrium solver \opt{EOS} \citep{2017A&A...597A..16P},
the continuous opacity package \opt{CONTOP}, the line opacity package \opt{LINOP} \citep{2015ascl.soft07008B},
and a plane-parallel/spherical radiative transfer solver based on Bezier-spline approximation to the source
functions \citep{2013ApJ...764...33D}.

The whole package is divided into two parts: the graphical user interface (GUI) that handles data input/output
and presentation/inspection of the results, and the actual solver that either generates the synthetic spectrum
or performs fitting of the observations. These two parts are communicating with each other through a data
structure, known as the SME structure, which contains all the information necessary for the calculations.
To initiate a process the user needs to create this structure either using the GUI or any other method. The
calculation part then uses the input structure to initiate its work and upon completion packages the results
into a similar output structure. The latter can explored with the GUI.

At the time of the initial development of SME, IDL was widely used in astronomy for data reduction and data
analysis. Therefore only the performance-critical parts of SME were written in C and FORTRAN. However, new
developments since then, such as the recent marketing policy of IDL, have motivated us to move SME to an
open and free platform, such as Python. This has the benefit that \pysme no longer relies on commercial
software, simplifies parallelization and opens additional functionality through the use of existing
Pyhthon libraries. This also simplifies the use of \pysme for large surveys as was done for example for the
GALAH \citep{2021MNRAS.506..150B} with the \idlsme. Additionally, the switch to Python offers us an
opportunity to improve and modernise several of components of SME. In the following sections we will
discuss in more detail specific changes that were made to \pysme.

\subsection{Optimization Algorithm}
\label{sec:algorithm}
One of the most useful capabilities of \pysme is its ability to to determine the best fit stellar parameters
from comparison with an observations. In \pysme this is done by solving the least squares problem:
\begin{equation}
    \label{eq:resid}
    \sum_i \text{weight}(\lambda_i)\cdot  r_i^2 = \mathrm{min}, 
\end{equation}
where the residuals $r_i\equiv\text{synth}(\lambda_i, p0, p1, ...) - \text{obs}(\lambda_i)$, $\text{synth}$
is the synthetic spectrum generated by \pysme, $\text{obs}$ is the observed spectrum, $p0, p1, ...$ are the
fitting parameters (e.g. effective temperature \steff), and $\lambda_i$ is the wavelength of point $i$.
$\text{weight}$ is the weight of each point conventionally set to the inverse of data uncertainties $\sigma$.
In \idlsme the weights where modified to include the residual intensity of observations: $\text{weight}_i^2=\text{obs}_i/\sigma_i^2$. In this way the optimisation algorithm is discouraged from
improving the fit to the line cores at the expense of the continuum points. We find this scheme useful,
in particular, when fitting a short spectral interval with comparable amount of continuum and line points
and so we kept this weighing scheme in the \pysme.

\begin{figure}
    \centering
    \includegraphics{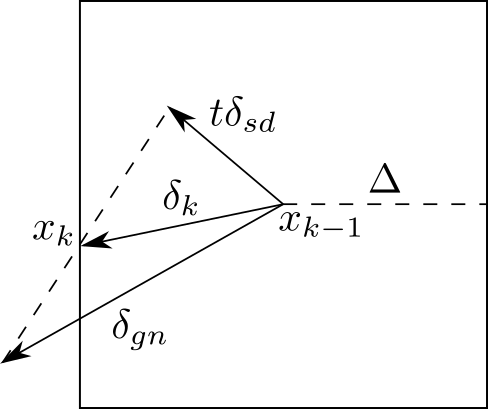}
    \caption{Sketch of the \opt{dogbox} algorithm in two dimensions for step $k$.
    The previous point is $x_{k-1}$, the next point is $x_k$. The rectangle
    represents the trust region with size $\Delta$, $\delta_{gn}$ is the
    Gauss-Newton step, and $t \delta_{sd}$ is the steepest decent step.
    Here the Gauss-Newton step falls outside the trust region, so the
    new step $\delta_k$ is a combination of both the Gauss-Newton step
    and the steepest decent, cut off at the boundary of the trust region.}
    \label{fig:dogbox}
\end{figure}

Instead of the Levenberg-Marquardt (LM) algorithm employed in the \idlsme, \pysme uses
the \opt{dogbox} algorithm as described in \citet{Voglis_arectangular} and implemented
in SciPy \citep{2020SciPy-NMeth}. This method is similar to the LM algorithm in that
it uses both the Gauss-Newton step and the steepest decent, but it uses an explicit
rectangular trust region. At each iteration it calculates the Gauss-Newton step and
the steepest decent, the next step is then one of the following three options:
\begin{itemize}
    \item If the Gauss-Newton step is within the trust region: Use the Gauss-Newton step
    \item If the Gauss-Newton step is outside the trust region, but the steepest decent step
    is inside it, use a combination of the two as shown in \autoref{fig:dogbox}.
    \item If both are outside the trust region: Follow the steepest decent direction to the
    edge of the trust region
\end{itemize}
The size of the trust region $\Delta$ is adjusted between iterations depending on the
improvement of the residuals. The size of the trust region is kept constant unless the
changes stop matching the linear estimate for the convergence. If the improvement is
larger than expected trust region is increased. If it is marginal the trust region
is decreased.

This method in our experience results in the least number of function evaluations
to reach the convergence out of the methods implemented in SciPy. It also allows us to set limits on the stellar parameters without confusing the
fitting process. This is occasionally required to prevent SME from extrapolating
outside the grid of stellar atmospheres and keep the interpolated model physical.

\subsection{Continuum Fitting}
\label{sec:continuum}
\begin{figure*}[ht]
    \centering
    \subfloat[\opt{mask+linear}]{\includegraphics[width=0.45\textwidth]{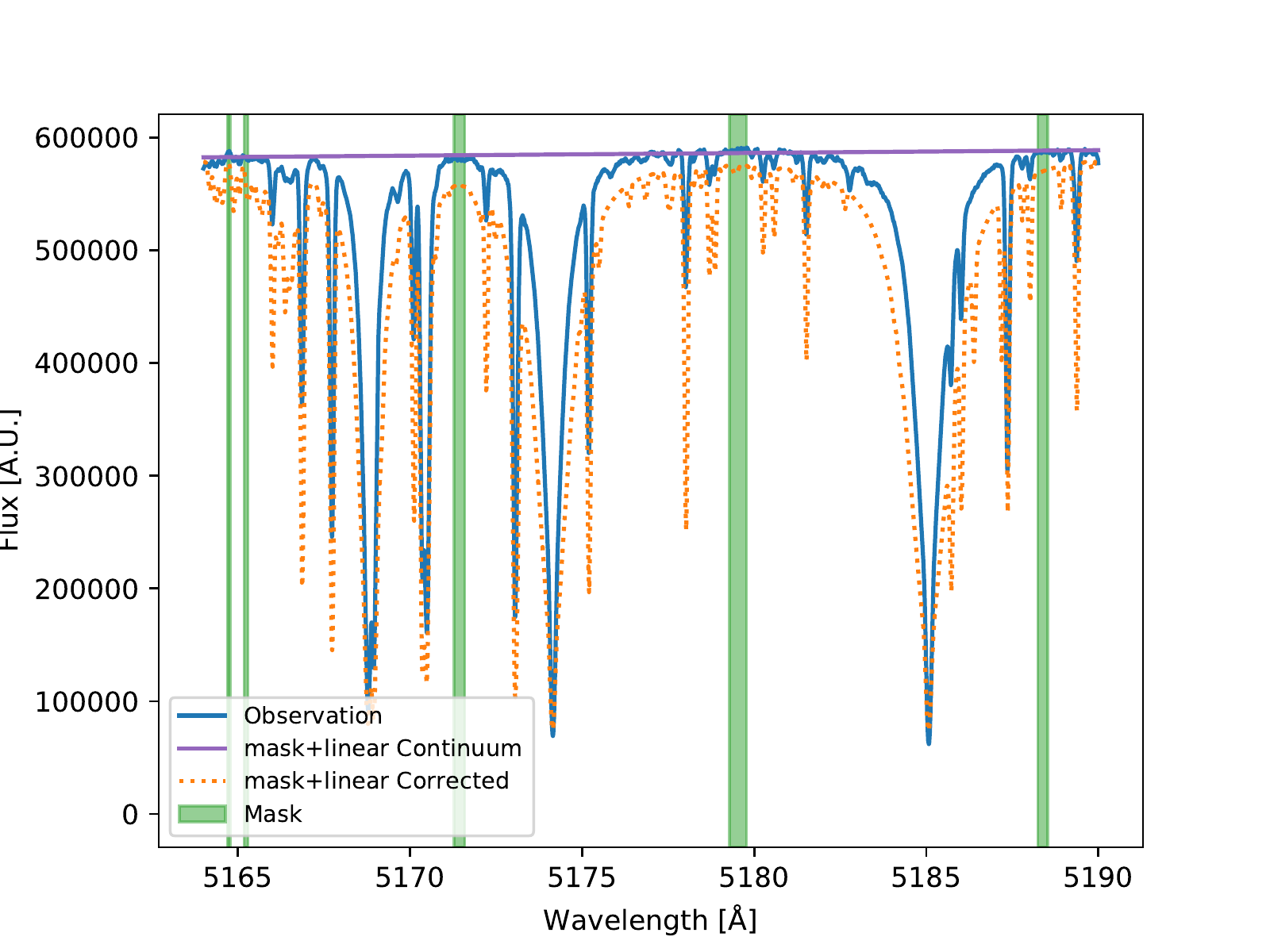}}
    \subfloat[\opt{mask+quadratic}]{\includegraphics[width=0.45\textwidth]{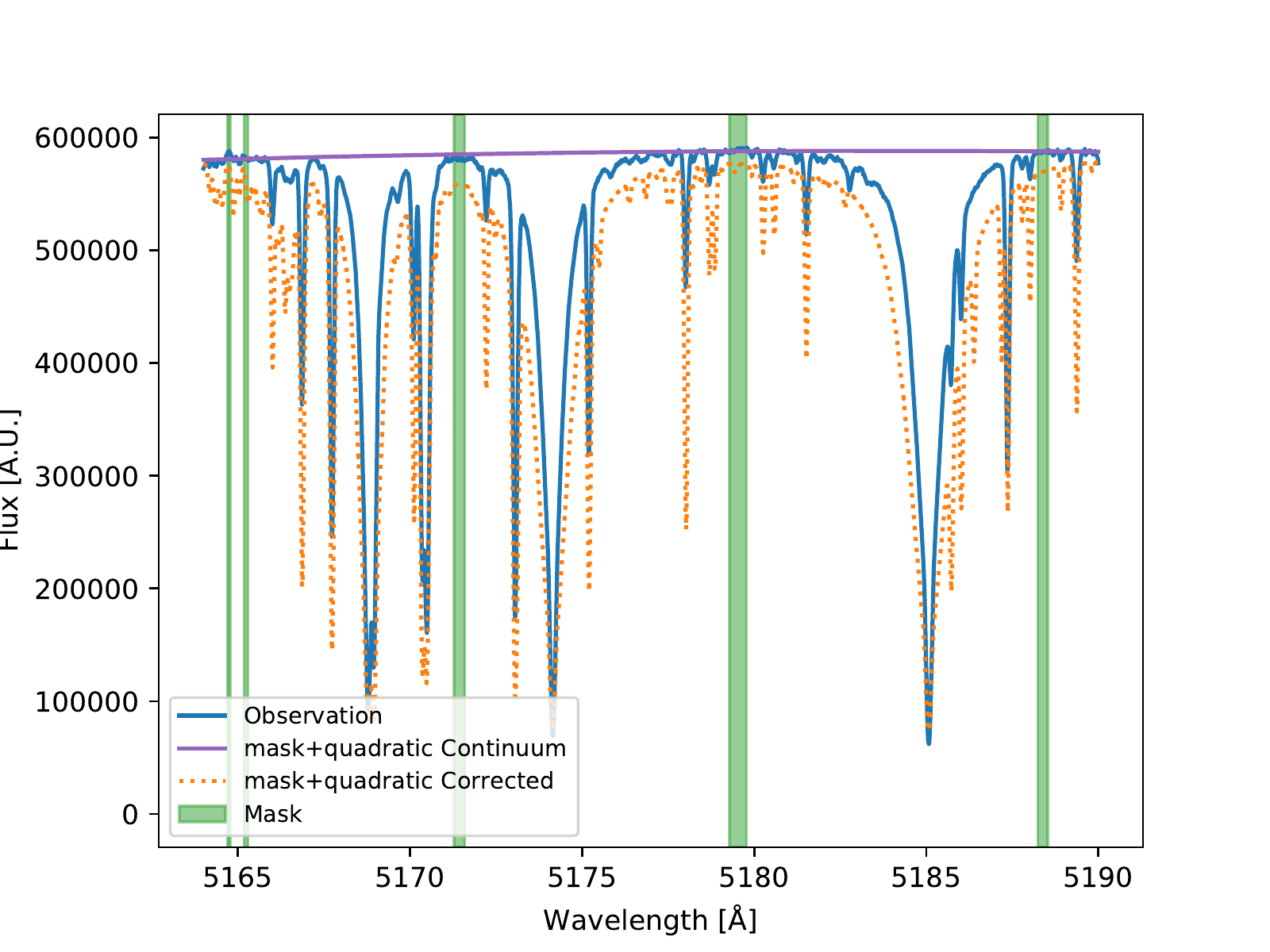}} \\
    \subfloat[\opt{match+linear}]{\includegraphics[width=0.45\textwidth]{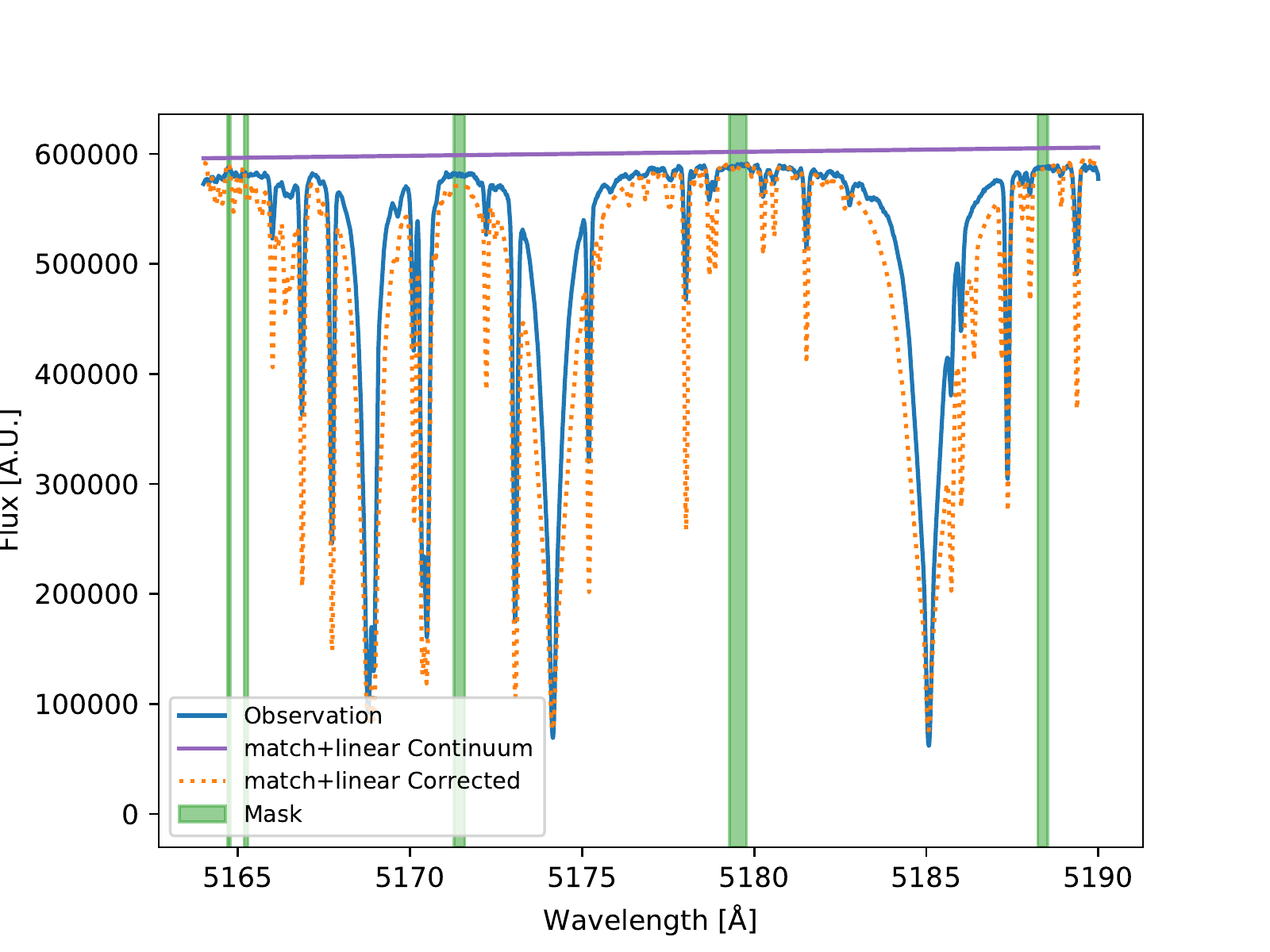}}
    \subfloat[\opt{match+quadratic}]{\includegraphics[width=0.45\textwidth]{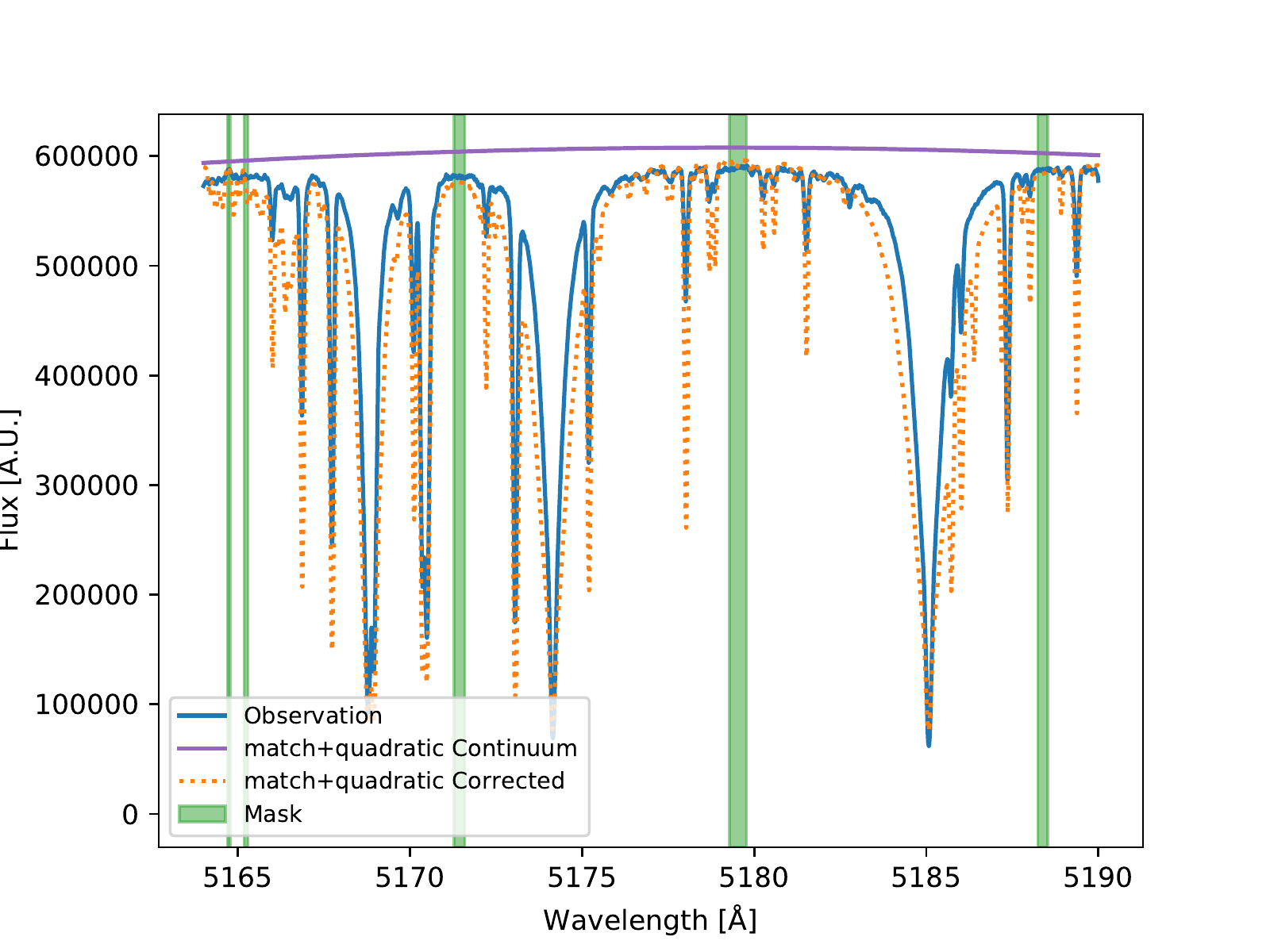}}
    \caption{Comparison of the different continuum normalization options on a small segment of the solar spectrum. Each plot shows the original observation (blue), the corrected synthetic spectrum (dotted orange), and the continuum detected by the method (purple). The shaded regions mark the selected pixels used as the mask (green).}
    \label{fig:continuum}
\end{figure*}

Correct continuum normalisation is very important for the accurate
determination of stellar parameters. In \pysme we achieve this by
changing the continuum of the synthetic spectrum to match the observed
spectrum, similar to \idlsme. Since every spectrum is different, there
is no good solution that fits all situations. We therefore implement
several continuum correction alternatives in \pysme. A list
of the methods is presented in \autoref{tab:continuum_options} and a
longer explanation can be found in \pysme documentation\footnote{\url{https://pysme-astro.readthedocs.io/en/latest/}}. Each continuum
fitting method has an associated degrees of freedom parameter, which
meaning depends on the selected method. Example of some options are
given in \autoref{fig:continuum}
for an F7 dwarf HD~148816 \citep{1999MSS...C05....0H} with
$\steff=\SI{5908}{\uteff}$,  $\slogg=\SI{4.32}{\slogg}$,
and $\smonh=\SI{-0.71}{\umonh}$ \citep{2020A&A...634A.136C}. 

In the \opt{mask} method a user-defined mask is used to specify
continuum points. Then a polynomial fit to those points is used
as the continuum of the spectrum. The degree of the polynomial
is a user-defined parameter. The benefit of this method is that
it allows good control over the continuum placing and it works
reliably if continuum points are present in selected wavelength
regions. Moreover, it does not rely on the synthetic spectrum,
so the continuum stays the same throughout SME iterations.
The downside is the need for interactive setting of the mask and
the requirement of having continuum points may be impossible
to meet, e.g. for TiO molecular bands in M-dwarfs or for regions
around strong lines with very broad wings.

An alternative approach is to match the relative depth of various
lines to the spectral synthesis. This is implemented as \opt{match}
option, which uses the fact that line depth is affected non-linearly
by the selected continuum level: spectral points in the center of
strong lines are much less affected by a change in continuum
placing than points in weak lines. Thus, the idea is to fit
the continuum so that:
\begin{equation}
    \label{eq:cont}
    \sum_i \left[\text{obs}(\lambda_i)-\text{syn}(\lambda_i) \cdot f(\lambda_i)\right]^2 = \mathrm{min},
\end{equation}
where $f$ is an analytical continuum function. We choose $f$ to be
a polynomial and determine the coefficients using a least-squares
fit. The degree of polynomial is a user-defined parameter. To
match the concept of continuum \pysme sets the weights for
spectral points proportional to their residual intensity,
so that a good continuum is found even if some lines are missing
in the synthesis. This method has the advantage that it does not
require the continuum points to be present in observations.
On the other hand it needs the observations to represent a good range
of line depths and a reasonable match to the synthetic spectrum.
This frequently means fitting a relatively large spectral ranges and works best for observations with a high signal-to-noise ratio (SNR).
Note, that this option requires re-evaluation of the continuum
on every SME iteration. The other methods listed in \autoref{tab:continuum_options} are various combinations of the \opt{mask} and the
\opt{match} methods.

Finally, \pysme can also rely on continuum normalization done
before the import of observations making no continuum adjustments. \autoref{fig:continuum} shows some of the more successful fits
to a spectral fragment for HD~148816.

\begin{table}[ht]
    \centering
    \begin{tabularx}{\columnwidth}{lX}
        \toprule
         Method & Description \\
         \midrule
         \opt{mask} & Polynomial fit to the selected points \\
         \opt{match} & Polynomial fit to match the synthetic \\
                      & and observed spectra \\
         \opt{match+mask} & Same as \opt{match}, but only uses mask points \\
         \opt{matchlines} & Similar to match, but it focuses on matching the line cores instead of the continuum \\
         \opt{matchlines+mask} & Same as \opt{matchlines}, but only uses mask points \\
         \opt{spline} & Similar to \opt{match}, but uses a cubic spline instead of a polynomial \\
         \opt{spline+mask} & Similar to \opt{match+mask}, but uses a cubic spline instead of a polynomial \\
         \bottomrule
    \end{tabularx}
    \caption{Continuum Normalization options}
    \label{tab:continuum_options}
\end{table}

\subsection{Radial Velocity Fitting}
\label{sec:rv}
Just as important as the continuum is the radial velocity shift
between the observed spectrum and the synthetic spectrum. The
radial velocity determination can be done for each wavelength
segment separately or for the entire spectrum. In \pysme we
determine the radial velocity in two steps. First we perform
a large-range search using the cross correlation between the
observed spectrum and the synthetic spectrum to get a rough
estimate for the radial velocity value and avoid shallow local
minima. The accuracy of this first guess is limited by the
wavelength resolution of the observation as we avoid¨
interpolation and perform cross correlation data pixels.
By default we limit the range of this search to \SI{\pm 500}{km/s},
but this value can be adjusted by the user. In a second step
we refine the radial velocity value in a least squares
sense, starting from the previous estimate.
We shift the synthetic spectrum wavelength using the relativistic
Doppler shift formula (\autoref{eq:rv}) in each iteration until
it matches the observed spectrum.

\begin{equation}
    \lambda^\prime = \lambda \sqrt{ \frac{1 + v / c}{1 - v / c}} \ , \label{eq:rv}
\end{equation}
where $\lambda^\prime$ is the wavelength shifted into the rest
frame of the star, $\lambda$ is the wavelength in the rest frame
of the observer, $v$ is the radial velocity (positive for a star
moving away from the observer), and $c$ is the speed of light.

\subsection{Auxiliary Data}
\label{sec:data}
The radiative transfer calculations at the core of SME require
additional data with atomic and molecular line properties (line
lists), as well as stellar model atmosphere(s). Additionally to
correct for Non Local Thermal Equilibrium (NLTE) effects, \pysme
requires NLTE departure coefficients matching the selected
atmospheric model. All of these are discussed in this section.

The line data for each transition must at least include the
species name, ionisation stage, excitation energy of the lower
level and the transition oscillator strength. These can be
complemented by the line broadening parameters for the natural,
Stark, and van der Waals broadening mechanisms. If they are not
known \pysme will use approximations as described in
\cite{1996A&AS..118..595V} and \cite{2017A&A...597A..16P}.
Finally for NLTE corrections (see below), the line list should
also include the term designation for the lower and upper
energy levels, as well as their angular momentum quantum
numbers J. Conveniently \pysme supports the linelist format
returned by VALD3
\citep{2015PhyS...90e4005R,2000BaltA...9..590K,1999A&AS..138..119K,1997BaltA...6..244R,1995A&AS..112..525P}
for the \opt{extract stellar} query, which includes all required
information. Both \opt{short} and \opt{long} formats are supported,
but the \opt{long} format is required for NLTE corrections.

Spectral synthesis in \pysme also needs a stellar atmosphere
model, which describes the temperature and pressure profiles
as functions of "depth", where depth could be an optical depth
at a standard wavelength or a column mass. When the effective
temperature, surface gravity or metallicity of the synthesis
does not match that of a pre-computed model, \pysme interpolates
in a 3-dimensional (\steff, \slogg, and \smonh) grid. Grids
for use with \pysme are provided by several authors. These
include the MARCS models \citep{2008A&A...486..951G} for cool
dwarf and giant stars, and ATLAS \citep{2017ascl.soft10017K,2002A&A...392..619H}
and LL models \citep{2004A&A...428..993S} for hotter main
sequence objects. Each model grid is packed in rather large
(on the order of GB) data file and multiple grids are often
available for different microturbulence parameters,
alpha-element abundances etc. 

To account for NLTE effects \pysme uses the departure
coefficient tables. Departure coefficient is the ratio
between the NLTE and the LTE population of an energy level
involved in radiative transition (\citealt{2017A&A...597A..16P}
describes the impact on absorption coefficient and the
source function). Departure coefficient depends on the
local physical conditions described by model atmosphere
and the abundance of the species, responsible for
absorption/emission. Thus, fitting specific abundance or/and
atmospheric parameters may require interpolation. For
that departure coefficients need to be computed for each
layer of every atmospheric model in a grid, and for several
elemental abundances around the model metallicity. The
resulting data files are even larger than the model grids,
and notably they can only be used with the model atmosphere
grid they were calculated for.

In the original \idlsme all these files where included with
the distribution package bringing its size to over 1 TB,
even though most users only need a subset of this data. In
\pysme atmosphere model grids and departure coefficient files
are instead stored on a data server until requested by the
user, at which point they are automatically downloaded. This
significantly reduces the installation footprint of \pysme in
comparison to \idlsme and, in addition this allows \pysme to
provide updates for these data files when available. Of course,
it is still possible to add custom grids.

The new default atmosphere grid is the \opt{marcs2014} grid,
which is essentially the \opt{marcs2012} grid with some holes
filled and with an improved spherical models
(T. Nordlander, priv. comm.). Additionally \pysme also supports
the latest NLTE departure coefficient grids by
\citet{2020A&A...642A..62A}, which have been calculated for the
2014 MARCS atmospheric grids.

\subsection{Elemental Abundances}
\label{sec:abund}
In addition to the overall metallicity, \pysme also allows setting
individual elemental abundances for the first \num{100} elements
(up to Einsteinium) manually or as a free parameter in the fitting.
\pysme supports abundance input following different conventions.
These include the "H=12" convention with elemental abundances set
relative to Hydrogen and the "SME" convention where abundances are set
relative to the total number of atoms in a volume.
Internally they are all converted to the \opt{H=12} format in the
Python part of \pysme, the SME library, however, uses the \opt{SME}
format, which was the only format used in \idlsme.

For convenience \pysme supports three alternatives for "default"
solar abundances, as described in \autoref{tab:solar_abundances}.
These replace the solar abundance values of \idlsme, which was
evolving with time.

\begin{table}[ht]
    \centering
    \begin{tabularx}{\columnwidth}{llX}
        \toprule
        Option & Source & Note \\
        \midrule
        \opt{asplund2009} & {\citet{2009ARA&A..47..481A}} & The default used by \pysme. \\
        \opt{grevesse2007} & {\citet{2007SSRv..130..105G}} & \\
        \opt{lodders2003} & {\citet{2003ApJ...591.1220L}} & \\
        \bottomrule
    \end{tabularx}
    \caption{Solar abundance sets included in \pysme.}
    \label{tab:solar_abundances}
\end{table}

\subsection{Convergence}
\label{sec:convergence}
\begin{figure}
    \centering
    \includegraphics[width=0.9\columnwidth]{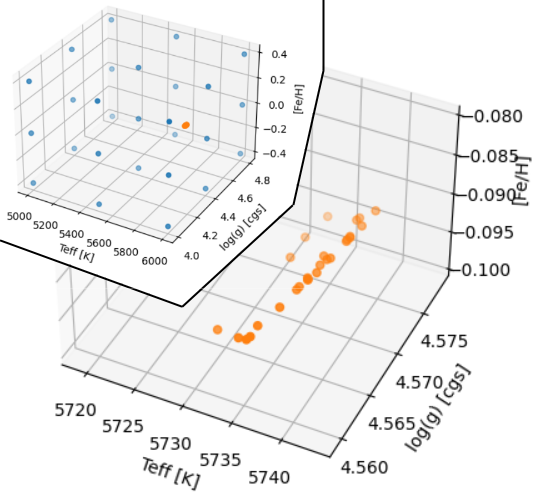}
    \caption{Distribution of the best fit stellar parameters \steff,
    \slogg, and \smonh for different initial parameters. Initial values
    are shown in blue and final values in orange. The cutout in the upper
    left corner shows the entire parameter space spanned by the initial
    parameters, while the rest is zoomed in on the closely clustered
    final distribution.}
    \label{fig:convergence}
\end{figure}

We test \pysme convergence to ensure the robustness of the new
optimisation algorithm, i.e. we test if we get similar results
for different initial sets of parameters. For this test we use
a segment of the solar spectrum (4489\,\AA\ to 4603\,\AA) 
provided by the National Solar Observatory Atlas 1 \citep{1984sfat.book.....K},
which is an optical flux spectrum for a relatively inactive Sun. We then determine the best fit \steff, \slogg, and \smonh, starting with different initial parameters on a 3x3x3 grid as given in \autoref{tab:convergence_input}.
The derived parameters of these 27 runs are all within the
uncertainties of the fit. In \autoref{fig:convergence}
we show the distribution of the points in the parameter space.
The standard deviation of the final values are $\sigma_{\steff} = 2.0$, $\sigma_{\slogg} = 0.005$, and $\sigma_{\smonh} = 0.001$.

\begin{table}[ht]
    \centering
    \begin{tabular}{lrrr}
        \toprule
        Parameter & Value1 & Value2 & Value3 \\
        \midrule
        \steff [\uteff] & 5000 & 5500 & 6000 \\
        \slogg [\ulogg] & 4.0 & 4.4 & 4.8 \\
        \smonh [\umonh] & -0.4 & 0.0 & 0.4 \\
        \bottomrule
    \end{tabular}
    \caption{Initial Parameters for the convergence test.}
    \label{tab:convergence_input}
\end{table}

\subsection{Parameter Uncertainties}
\label{sec:uncertainties}
As in all parameter determinations there are two different types of uncertainties to measure. The first is the statistical uncertainty, which depends on the SNR of the observations and is easily determined from the least squares fit in \pysme using the covariance matrix. We correct these uncertainties in \pysme using the final $\chi^2$, by multiplying the covariance matrix with $\sqrt{\chi^2}$. This normalized the initial uncertainties on the data points to those expected for this fit.

The second uncertainty is the systematic uncertainty, which is due to inherent deficiencies of the atomic and molecular data as well as observational defects (e.g. due to incorrect parameters in the line list). These are a lot more difficult to constrain, as there is no reference for \pysme to use. Instead \pysme uses the uncertainty method described in \citet{2016MNRAS.456.1221R,2017A&A...597A..16P}. This method uses the fit residuals, derivatives, and uncertainties of the observed spectrum to determine the cumulative probability distribution under the assumption that the entire residual can be explained by the variation in one parameter. Ignoring the sensitivity to other free parameters leads to an overestimation of the uncertainties. This approach works reasonable well for free parameters that explicitly affect the majority of spectral points (e.g. \steff, \smonh) but the estimate becomes unrealistically exaggerated for parameters affecting spectra locally (\slogg, individual abundances). We also note here that this method is invariant to the absolute scale of the input uncertainties of the spectrum.

\begin{figure}[!ht]
    \centering
    \subfloat[\steff]{\includegraphics[width=0.9\columnwidth]{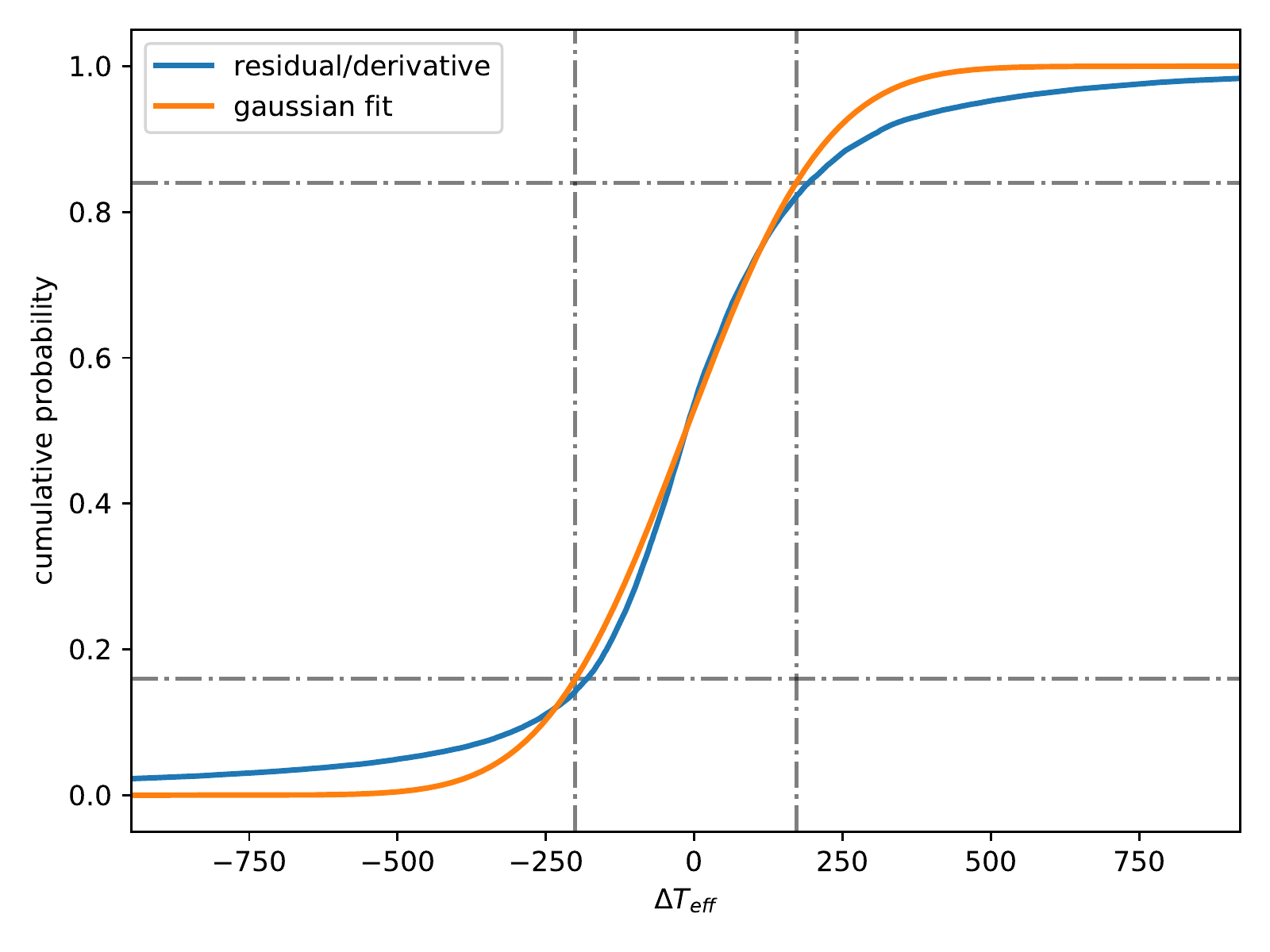}} \\
    \subfloat[\steff]{\includegraphics[width=0.9\columnwidth]{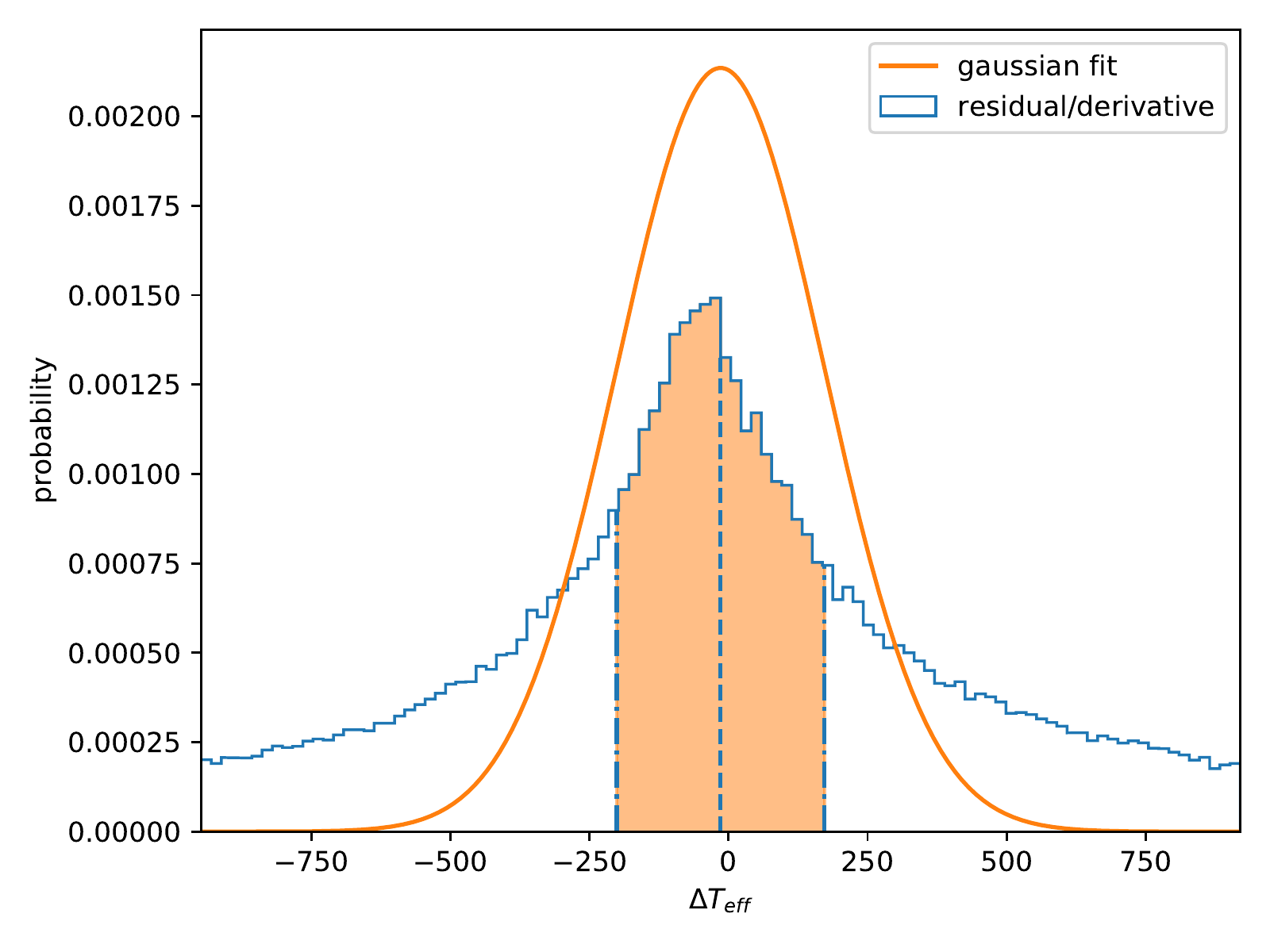}}
    \caption{Cumulative probability (top) and probability density (bottom) distribution of
    \steff\ for a spectrum of \epseri derived from the fit results (blue). For comparison we
    also show the distribution of a Gaussian with the same median and $\sigma$ as derived
    from the cumulative probability (orange). The median is marked by a dashed line (bottom), and
    the $1\sigma$~range is shown as dash dotted lines and a shaded region (bottom).
    The distributions for the other fitted stellar parameters are shown in the appendix
    \autoref{fig:cum_prob_logg_app}}.
    \label{fig:eps_eri_cum_prob_teff}
\end{figure}

In \autoref{fig:eps_eri_cum_prob_teff} we show the distribution for uncertainties for \steff\ for \epseri (analysis details are discussed in \autoref{sec:analysis}) as cumulative probability and as probability density\footnote{Distributions for the other parameters can be found in \autoref{sec:uncertainties_appendix}}. Note that the central part looks reasonably close to a Gaussian distribution, which explains why we get realistic estimate for the uncertainty of \steff. For other, more local parameters, such as individual abundances, the central part is often very asymmetric and quite different from a Gaussian.

\begin{figure}[ht]
    \centering
    \includegraphics[width=\columnwidth]{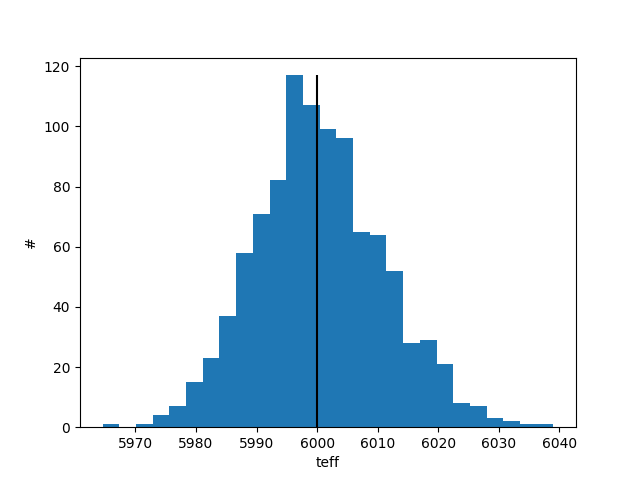}
    \caption{Scatter of the recovered \steff around the true value for a synthetic spectra with random noise of SNR \num{100}.}
    \label{fig:uncs_mc}
\end{figure}

To evaluate the accuracy of these uncertainty estimates we run a simple Monte-Carlo test. For this we create a synthetic spectrum for a single wavelength segment between \SIrange{6400}{6500}{\AA} with stellar parameters $\steff = \SI{6000}{\uteff}$, $\slogg = \SI{4.4}{\ulogg}$, $\smonh = \SI{0}{\umonh}$. Then we apply white noise with a SNR of \num{100} to this spectrum and extract the stellar parameters \steff, \slogg, and \smonh using \pysme with the initial parameters disturbed around the true values. Repeating this process for \num{1000} we can estimate the uncertainty of the parameters from the scatter of the values. This scatter is shown in \autoref{fig:uncs_mc} for \steff, the other parameters show a similar behaviour. A comparison between the values is given in \autoref{tab:uncs_mc}. This process does not include the systematic uncertainties of the input data as we create a synthetic spectrum to compare to, thus the uncertainties derived from the scatter match those derived by the least squares fit.

A final way to estimate the uncertainties is by comparing the results obtained here with those from previous studies (see \autoref{sec:trends}). Using the scatter of the differences we can then infer an estimate for the systematic uncertainties. The systematic uncertainties obtained this way are: $\sigma_{\steff} = 121$, $\sigma_{\slogg} = 0.16$, and $\sigma_{\smonh} = 0.07$.

\begin{table}[ht]
    \centering
    \begin{tabular}{lrrrr}
         \toprule
         Parameter & $\sigma_\text{mc}$ & $\sigma_\text{fit}$ & $\sigma_\text{sme}$  & $\sigma_\text{delta}$\\
         \midrule
         \steff [\uteff] & \num{10.2} & \num{14.5} & \num{54} & \num{121}\\
         \slogg [\ulogg] & \num{0.020} & \num{0.027} & \num{0.28} & \num{0.16}\\
         \smonh \umonh & \num{0.0065} & \num{0.0081} & \num{0.04} & \num{0.07}\\
         \bottomrule
    \end{tabular}
    \caption{Comparison between the uncertainty estimates based on a synthetic spectrum. $\sigma_\text{mc}$ is based on the distribution of a Monte-Carlo simulation, $\sigma_\text{fit}$ is based on the covariance matrix of the least-squares fit, $\sigma_\text{sme}$ is based on the method described in \autoref{sec:uncertainties}, and $\sigma_\text{delta}$ is based on the scatter of the derived values compared to the reference values from other studies in \autoref{sec:trends}.}
    \label{tab:uncs_mc}
\end{table}

\subsection{Graphical User Interface}
\label{sec:gui}
For \pysme an entirely new graphical user interface (GUI) was
developed\footnote{available here \url{https://github.com/AWehrhahn/PySME-GUI/}}.
For improved usability it is relying on established web technologies using the Electron\footnote{\url{https://www.electronjs.org/}} framework.
The interface is divided into a number of sections, each managing
one part of the SME structure. The spectrum section allows the user
to zoom and pan on the spectra both observed and synthetic, together
with the positions of the lines in the linelist. Furthermore it is
possible to manually set and manipulate the bad pixel and continuum
normalization mask.

The other sections can be used to change the parameters of the SME
structure, including the entirety of the linelist, the elemental
abundances, and the NLTE settings. 

It does however not offer some of the functionality that the \idlsme
GUI included. For example it is not possible to select lines directly
in the spectrum, to inspect or modify their parameters. Neither is it
not possible to measure the equivalent width of lines directly in the
interface.

\subsection{Interopability between \idlsme and \pysme}
\pysme has been designed for an easy transition for existing \idlsme
users. It is therefore capable of importing existing \idlsme input
(\opt{.inp}) and output (\opt{.out}) structures for use with \pysme or
its GUI. \pysme itself however uses a new file format to store those
structures (\opt{.sme}), which can currently not be used by \idlsme.
It is possible though to create IDL readable files if an IDL
installation is available on the machine using the \opt{save\_as\_idl}
function.

\subsection{Parallelization}
In some circumstances users may be interested in analyzing a large number
of stellar spectra, e.g. in surveys. Previously this required an
individual IDL license for each process that is running at the same time.
This is expensive (IDL cluster license) and complicated. \pysme solves
both of these problems at once. Still one has to be careful, it is
recommended to run each synthesis or fitting in a separate process, 
since the SME library should not be shared between them. Many tools
exist for different systems, for example the GNU parallel tool
\citep{GNUPARALLEL}. \pysme includes an easy example of such a script.

\subsection{Open Source}
In addition to replacing IDL with Python, which is Open Source, we also made the SME library available under the BSD 3-Clause
license, which is an Open Source license. This means that you can use \pysme in your projects which specify open source
requirements as part of your funding, which is in line with the European Commission's Open Science guidelines.

\section{Comparison between \idlsme and \pysme}
\label{sec:comparison}

With a new version of SME it is interesting to compare \pysme to \idlsme,
which we here split into two parts. First we assure that both versions
reach the same set of free parameters (within the estimated uncertainties)
given the same spectral synthesis as observations. In a second step we compare the
performance, i.e. the runtime, between the two version.

\subsection{Comparison of spectral synthesis}
\label{sec:comparison_resuts}
\begin{figure}[ht]
    \centering
    \includegraphics[width=1.\columnwidth]{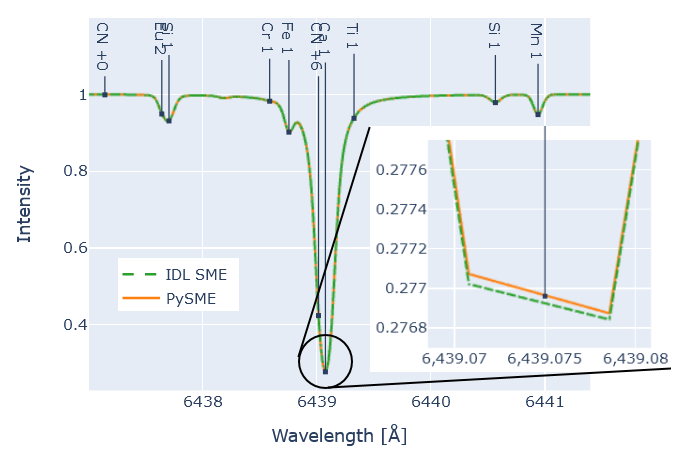}
    \caption{Comparison between the \idlsme stellar synthesis (dashed green)
    and the \pysme stellar synthesis (orange), for a small section of the
    stellar spectrum. All lines in the line list are marked and labeled
    accordingly by vertical lines. The zoom in on the Fe line highlights
    the small numerical difference between the two codes.}
    \label{fig:compare}
\end{figure}

The results of the stellar synthesis are plotted in \autoref{fig:compare}.
At first glance the two codes seem to create identical synthetic spectra,
however upon closer inspection there are small numerical differences
as shown in the zoom in panel for the Fe line in \autoref{fig:compare}.
These differences are on the order of \SI{0.01}{\%} matching the declared
precision of the radiative transfer solver included in the library. That
these spectra differ by this amount is not trivial, as the Python/IDL layer of
SME performs significant calculations, including the interpolation in the
stellar atmosphere grid, wavelength re-sampling, disk integration with the
application of macro turbulence and rotational broadening as well as the
instrumental profile. We believe that the two implementations are
sufficiently close as not to cause additional change that exceeds the
precision of the radiative transfer solving.

To further convince ourselves that the new incarnation of SME performs spectral
fitting at least equally well as the \idlsme we have done a blind
comparison between the two. For that we used HARPS spectra of a K dwarf \cnc.
The \idlsme analysis was carried out by Ryabchikova (TR) while \pysme
fitting was done by Wehrhahn (AW). The only common parts in the two were the
observations, the spectral range and the SME library. More details
on the comparison and the results are presented later in \autoref{sec:targets}.

\subsection{Comparison of performance}
The second test concerns the speed of the calculation. For this purpose we
compare the execution time for the same short spectral interval introduced in
\autoref{sec:comparison_results} for both \pysme and \idlsme. Even on the same
machine the execution time varies due I/O performance and so we recorded the
minimum and the average time of \num{1000} runs presented in
\autoref{tab:comparison_performance}. This shows that the execution times between
\pysme and \idlsme are similar. Further analysis shows that the vast majority
of CPU time is spent in radiative transfer calculation in the SME library, which
is shared between both versions of SME.

\begin{table}[ht]
    \centering
    \begin{tabular}{lrr}
        \toprule
         & \pysme & \idlsme \\
         \midrule
         Minimum Time & \SI{0.53}{s} & \SI{0.48}{s} \\
         Average Time & \SI{0.75 \pm 0.05}{s} & \SI{0.72 \pm 0.05}{s} \\
         \bottomrule
    \end{tabular}
    \caption{Minimum and Average runtime of \pysme and \idlsme on the same machine
    over 1000 runs. The machine runs Ubuntu and has a \SI{3.60}{GHz} CPU.}
    \label{tab:comparison_performance}
\end{table}

\section{Analysis}
\label{sec:analysis}
Finally we apply \pysme to a small selection of targets. The target selection is discussed
in \autoref{sec:target_selection}, followed by discussions on the chosen settings (\autoref{sec:setup})
and a short discussion of each target (\autoref{sec:targets}). The final parameters are given in
\autoref{tab:hd22049}. The results are compared to each other in \autoref{sec:trends}.

\subsection{Target Selection}
\label{sec:target_selection}
Determination of accurate stellar parameters for a small number of exoplanet host stars is
important for interpretation of transit spectroscopy but also gives us an opportunity to
assess \pysme performance for a range of spectral types. Close proximity to the Sun and
presence of planets makes these stars interesting and so we expected to find additional
data, such as accurate parallaxes and interferometry, to help us set independent constraints
on some of the stellar parameters. The selection of targets include 9 stars from the
VLT CRIRES+ transit spectroscopy survey. High-quality spectra for these targets are available
from ESO archive (3.6m HARPS) while interferometric radii were taken from 
\citet{2017ApJ...836...77Y,2014MNRAS.438.2413V}. \autoref{tab:datasets} summarises the data
we used for the analysis.

\begin{table*}[ht]
    \centering
    \begin{tabular}{lrrrr}
         \toprule
         Star & Program ID & Archive ID & S/N & Interferometry\\
         \midrule
         \epseri  & 60.A-9036(A) & ADP.2014-10-02T10:02:04.297 & 374 & yes \\
         \hnpeg & 192.C-0224(A) & ADP.2014-10-06T10:04:55.960 & 149 & yes \\
         \hdten  & 083.C-0794(A) & ADP.2014-09-23T11:00:40.757 & 167 & yes \\
         \hdthirteen & 072.C-0488(E) & ADP.2014-10-01T10:22:40.410 & 125 & no \\
         \hdseventeen & 072.C-0488(E) & ADP.2014-10-02T10:00:41.887 & 200 & no \\
         \hdeighteen & 072.C-0488(E) & ADP.2014-09-16T11:05:45.457 & 158 & yes \\
         \cnc & 288.C-5010(A) & ADP.2014-09-26T16:51:14.897 & 135 & yes \\
         \wasp & 0104.C-0849(A) & ADP.2019-11-16T01:15:37.789 & 81 & no \\
         \bottomrule
    \end{tabular}
    \caption{List of stars analysed with \pysme including the ESO program and archive IDs, average
    signal-to-noise ratio and the availability of reliable interferometric measurements.}
    \label{tab:datasets}
\end{table*}

\subsection{Preparation of our test sample}
\label{sec:setup}

\subsubsection{Continuum}
\label{sec:continuum2}
The standard HARPS pipeline does not do continuum normalisation (not required for radial velocity measurements)
and thus our first task was to correct for the spectrometer blaze function and rely on \pysme correction for the
fine tuning as described earlier in \autoref{sec:continuum}. We determine the upper envelope of the spectrum by
selecting the local maxima and fitting a smooth function. The maxima are found by comparing neighbouring points,
and then only keep the largest local maximum in a given interval (step size). The step size should be larger than
the width of absorption features, but small enough to follow the continuum. For our data we choose the step size
to be \num{1000} pixels. Then we connect the selected points with straight lines and smooth the resulting curve
using a Gaussian of the same width as the step size. This creates a continuous and smooth fit, that is good enough
for further analysis.

The observations are then split into segments corresponding to \num{100000} pixels each. This number is arbitrary,
mostly set by easiness of inspecting the results.

When running \pysme the selected linear correction of the continuum correction based on best match to the synthetic
spectrum.

\subsubsection{Radial Velocities}
\label{sec:radial_velocities}
For the radial velocities we could rely on HARPS wavelength calibration and so we used a single radial velocity value for all
spectral intervals.

\subsubsection{Uncertainties}
\label{sec:analysis_uncertainties}
SME uses uncertainties of spectral points to compute the weights for the fitting procedure. The HARPS data does not
contain an independent estimate of uncertainties. Only the flux value in each pixel is available. We therefore assume that the Poissonian shot noise is the only source of noise. The data has a signal-to-noise ratio on the order (or in excess) of \num{100} making this a good assumption. We additionally add weight to the line centers by multiplying these uncertainties by the flux values. The final uncertainties are then: 
\begin{equation}
    \sigma = \sqrt{F_n}\ F_n \ ,
\end{equation}
where $F_n$ is the normalized flux of the observation.

\subsubsection{Tellurics}
\label{sec:tellurics}
As with any ground based observation our spectra also contain telluric absorption features. We therefore provide \pysme with a telluric spectrum generated with TAPAS \citep{2014A&A...564A..46B}, ignoring the Rayleigh scattering as it affects the continuum on much larger wavelength scales and will therefore be removed as part of the continuum correction. We also remove slow variations in the tellurics using the same method as for the spectrum as described in \autoref{sec:continuum2}. The telluric spectrum is used to mask any significant telluric lines in the observation. For this any points in the spectrum are marked as bad pixels if the tellurics are larger than \num{0.5}~\% of the normalized spectrum.

\subsubsection{Instrumental Broadening}
\label{sec:ipres}
La Silla HARPS has a slightly asymmetric instrumental profile but for our purposes deviations from a Gaussian are small and we use Gaussian broadening with a resolution of \num{105000} to match spectral synthesis with HARPS spectra.

\subsubsection{Stellar Disk Integration}
\label{sec:mu}
Disk integration combines the rotational broadening and broadening due to radial-tangential macroturbulence. As described in section 3.3 of the original paper (\citealp{1996A&AS..118..595V}) disk integration is carried out using quadratures. For the calculations described here stellar flux spectra are computed using specific intensity at \num{7} nodes (limb distances) allowing to reach the precision better than \SI{0.1}{\%}, typically $3\cdot10^{-4}$. Both SME implementations allow to adjust the number of nodes for better precision or faster computations while the nodes and weights are generated automatically.

\subsubsection{Line list} 
\label{sec:linelist}
We generate the line lists for all stars using VALD \citep{1995A&AS..112..525P}, covering the wavelength range from \SI{3781.22}{\AA} to \SI{6912.21}{\AA}. We use the same list for all stars, which includes \num{52496} individual lines with an expected depth of at least \num{1}~\% at \steff \SI{5770}{\uteff}, \slogg \SI{4.4}{\ulogg}, \smonh \num{0}, and \svmic \SI{1}{\uvmic}. 
All lines are given in the \opt{long} format necessary for NLTE corrections. The line list wavelengths are in air: even though the HARPS instrument works inside a vacuum chamber, since the data reduction pipeline converts the wavelength scale of the reduced spectra to air. All references for the line parameters in each element are given in \autoref{sec:linelist_references}.

\subsubsection{Model Atmosphere} 
\label{sec:model_atmosphere}
For the model atmosphere grid we chose the \opt{marcs2012} grid \citep{2008A&A...486..951G}, which is included in the \pysme distribution (see \autoref{sec:data}). We chose this grid, since it spans the parameter space of our target stars and supports the NLTE departure coefficient grids.

\subsubsection{Elemental abundances}
\label{sec:metallicity}
While we do fit the overall metallicity of our stars, we do not fit abundances of individual elements (even though \pysme is capable of doing so). The relative abundances are instead assumed to be solar as defined by \citet{2009ARA&A..47..481A}.

\subsubsection{NLTE departure coefficients}
\label{sec:nlte2}
We allow \pysme to apply NLTE corrections for the \num{13} elements: H, Li, C, N, O, Na, Mg, Al, Si, K, Ca, Mn, Ba using the departure coefficient grids described in \citet{2020A&A...642A..62A}. In addition, Fe NLTE corrections the departure coefficients Fe described in \citet{2016MNRAS.463.1518A}. All references for the NLTE grids are given in \autoref{sec:nlte_references}.

\subsubsection{Initial stellar parameters}
\label{sec:initial_parameters}
The least squares fit requires an initial guess of the stellar parameters. The better the guess is the less
iterations to determine the optimal fit parameters. The results may vary slightly depending on the of initial
but, as was shown in \autoref{sec:convergence}, our algorithm is quite robust and the space we explore
does not contain important local minima. Thus we opted to select stellar parameters from previous studies
given in the NASA Exoplanet Archive (NEXA, \citealt{nea13}) as the initial guess.
The comparison of our results with NEXA parameters and alternative estimates, based on interferometric
data together with the references is presented in \autoref{tab:hd22049}.

\subsection{Individual Targets}
\label{sec:targets}

\begin{figure*}
    \centering
    \includegraphics[width=\textwidth]{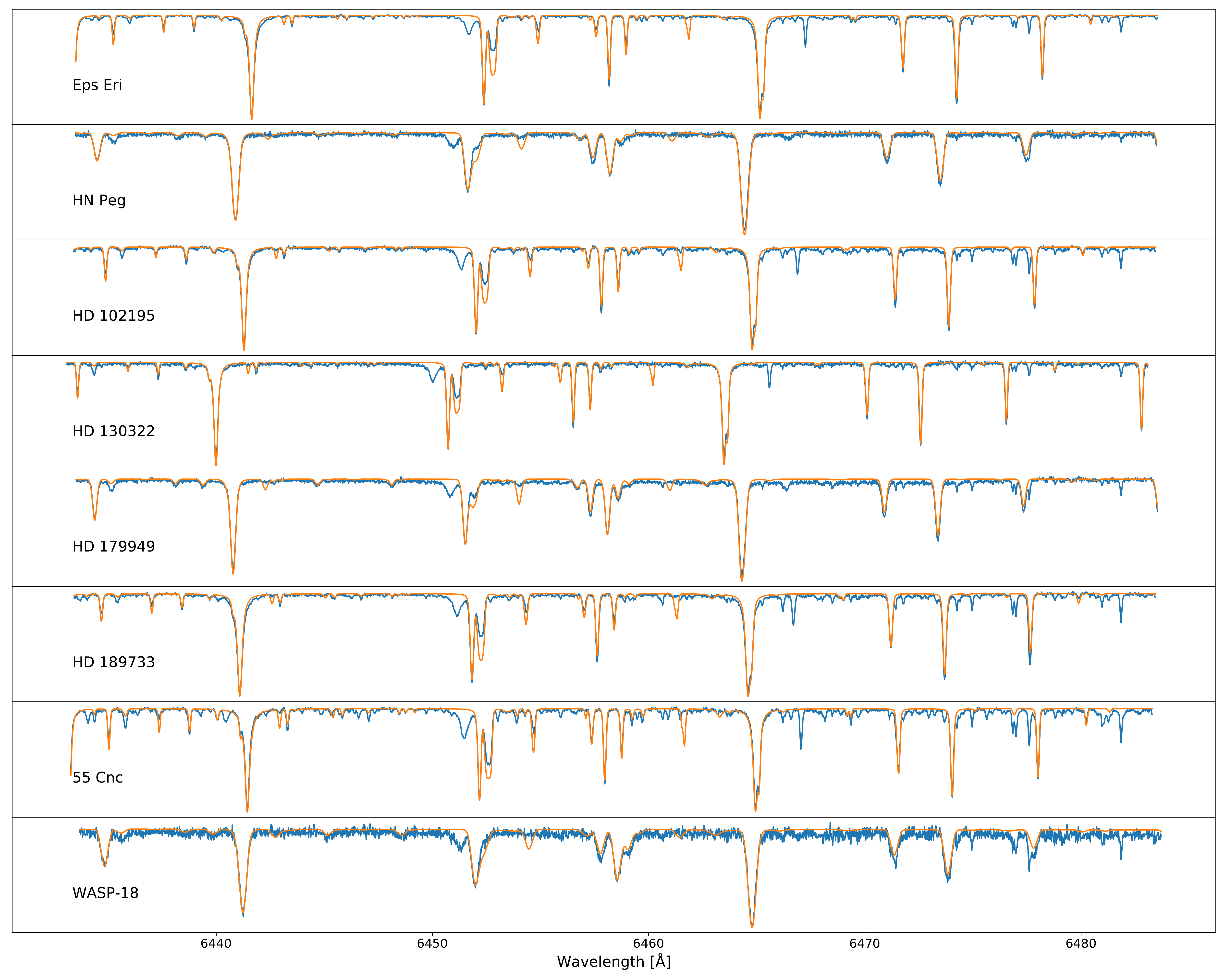}
    \caption{Observation (blue) and Synthesis (orange) flux of one wavelength section for all target stars.}
    \label{fig:plot_all}
\end{figure*}

We run \pysme on all our targets and compare the results with \idlsme (see \cnc\ below) and with other techniques (\autoref{sec:comparison_results}). The free parameters were \steff, \slogg, \smonh, \svsini, \svmic, \svmac
and \srv. \autoref{fig:plot_all} shows the final fit to the observations
for small but representative fragment of the spectrum.

\noindent\epseri
is a nearby sun-like star of spectral type K2 \citep{1989ApJS...71..245K}, with at least one planet \citep{2019AJ....157...33M,2006AJ....132.2206B,2000ApJ...544L.145H} and a proposed second planet
\citep{2002ApJ...578L.149Q}. \epseri\ is one of the Gaia FGK benchmark star \citep{2018RNAAS...2..152J}
resulting the the following parameters: \steff=5076K, , \slogg=4.61, and \smonh=-0.09. These are to be
compared with \pysme results: 4934K, 4.38, and -0.213. The HARPS data has the highest S/N in our sample
so we used the example of \epseri to illustrate the uncertainty estimation by \pysme in \autoref{sec:uncertainties}.\\

\noindent\hnpeg is a star of spectral type G0 \citep{2001AJ....121.2148G}, with one known exoplanet \citep{2007ApJ...654..570L}.
Of the stars in this paper it is most similar to the Sun.\\

\noindent\hdten is a K0 star \citep{1999MSS...C05....0H}, with one known exoplanet \citep{2006ApJ...648..683G}.\\

\noindent\hdthirteen is another K0 star \citep{1999MSS...C05....0H}, with one known exoplanet \citep{2000A&A...356..590U}.\\

\noindent\hdseventeen is an F8 star \citep{1988mcts.book.....H}, with one known exoplanet \citep{2001ApJ...551..507T}.\\

\noindent\hdeighteen is a K2 star \citep{2003AJ....126.2048G} in a binary system, with one known exoplanet
\citep{2005A&A...444L..15B}.\\

\noindent\cnc is a binary system with a G8-K0 dwarf (\cnc~A, \citealt{2003AJ....126.2048G}) and an M4 dwarf (\cnc~B,
\citealt{2015A&A...577A.128A}). Here we investigate \cnc~A that is a host to a complex system of five planets
\citep{2018A&A...619A...1B}. This star was independently analysed with \idlsme and so we used it the comparison between the two implementations of SME.

Observational data included several spectra of \cnc obtained with the La Silla HARPS instrument on January 2012. The raw data was reduced using the standard ESO HARPS pipeline and combined to produce a single spectrum with the mean S/N of 135. We (TR) started the \idlsme analysis by manually adjusting the continuum and selecting five spectral intervals: 4900-5500\,\AA\AA, 5500-5700\,\AA\AA, 5700-6000\,\AA\AA, 6000-6300\,\AA\AA, and 6300-6700\,\AA\AA. After computing a spectral synthesis based on stellar parameters taken from the literature (\cite{2000A&A...356..590U}) and the VALD3 line data we have adjusted the mask of "bad" pixels and let \idlsme do the final linear correction of the continuum level in each interval. We then solved for \steff, \slogg, \smonh, \svsini, and \srv. We assumed fixed \svmic and a Gaussian instrumental profile corresponding to the resolving power of HARPS.

After this step we revisited the mask to remove poorly fitted spectral lines with obviously erroneous line data. The decision was based on comparison with the average fit quality for lines of the same species. The final step was to re-fit the parameters listed above enabling NLTE correction for Fe, Mg, Mn, Na, Si, Ba and Ca.

\begin{figure*}
    \centering
    \includegraphics[width=0.9\textwidth]{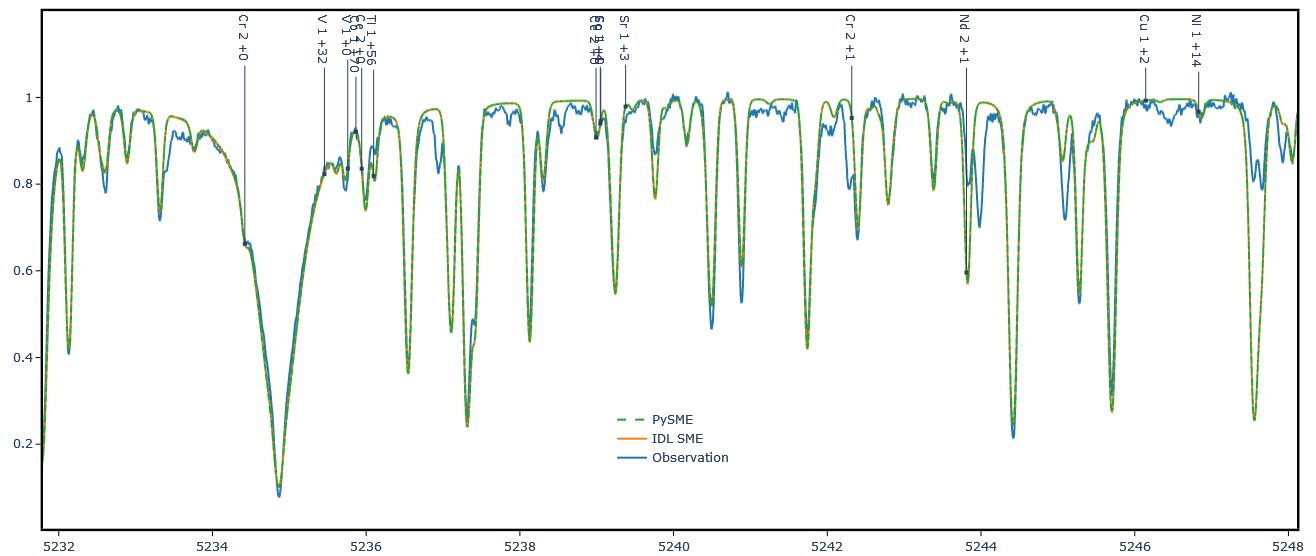}
    \caption{Comparison of the least squares fit results for \cnc between the \idlsme and the \pysme analysis. Both analyses used the same continuum normalization, mask, linelist, atmosphere grid, and NLTE departure coefficients. The difference between \pysme and \idlsme is small, but noticeable in the derived parameters.}
    \label{fig:comparison_lsq}
\end{figure*}

The \pysme analysis for the comparison used the same observational data, spectral intervals, continuum normalization, mask, linelist, atmosphere grid, and NLTE departure coefficients. Thus using the same input parameters, and only comparing the results of the fitting procedure. The resulting parameters are given in \autoref{tab:compare_55cnc} and \autoref{fig:comparison_lsq} shows the comparison for a fragment of spectral data included in the analysis. The difference between the best fit spectra is small and the derived parameters are compatible with each other, though not identical.

The values given in \autoref{tab:hd22049} are taken from an independent \pysme analysis that follows the same steps as all other stars, i.e. has a different continuum normalization, linelist, mask, and NLTE departure coefficients than the comparison analysis.

\begin{table}
    \centering
    \begin{tabular}{lrr}
         \toprule
         Parameter &  \pysme & \idlsme \\
         \midrule
         \steff [\uteff] & \num{5172 \pm 4} &  \num{5205 \pm 30} \\
         \slogg [\ulogg] & \num{4.17 \pm 0.007} & \num{4.26 \pm 0.09} \\
         \smonh \umonh & \num{0.30 \pm 0.003} & \num{0.36 \pm 0.05} \\
         \svmic [\uvmic] & \num{0.8} & \num{0.8} \\
         \svmac [\uvmac] & \num{3.4} & \num{3.4} \\
         \svsini [\uvsini] & \num{0.1} & \num{0.1} \\
         \bottomrule
    \end{tabular}
    \caption{Stellar parameters derived from the same HARPS spectrum of \cnc, in two independent analyses, one using \pysme and one using \idlsme. The \pysme analysis here uses different continuum normalization, linelist, mask, and NLTE departure coefficients than the analysis results shown in \autoref{tab:hd22049}.}
    \label{tab:compare_55cnc}
\end{table}

\noindent\wasp is an F6 star \citep{1978mcts.book.....H} in a binary system, with two known exoplanets \citep{2009Natur.460.1098H,2019AJ....158..243P}. This star has the lowest SNR of the sample with a SNR of \num{81}.\\


\section{Comparison with other studies}
\label{sec:comparison_results}

\begin{figure*}[ht]
    \centering
    \subfloat[\steff]{\includegraphics[width=0.45\textwidth]{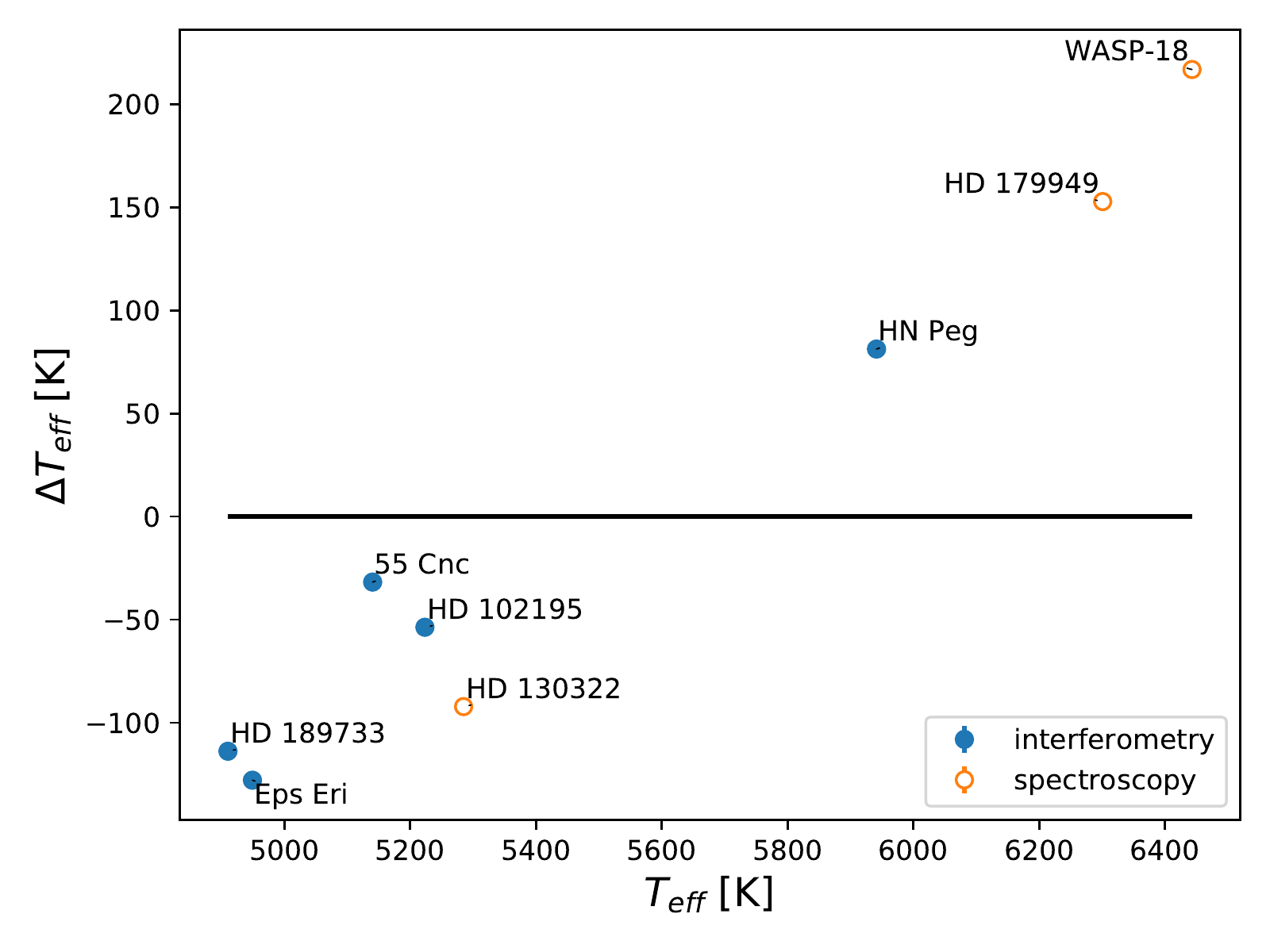}}
    \subfloat[\slogg]{\includegraphics[width=0.45\textwidth]{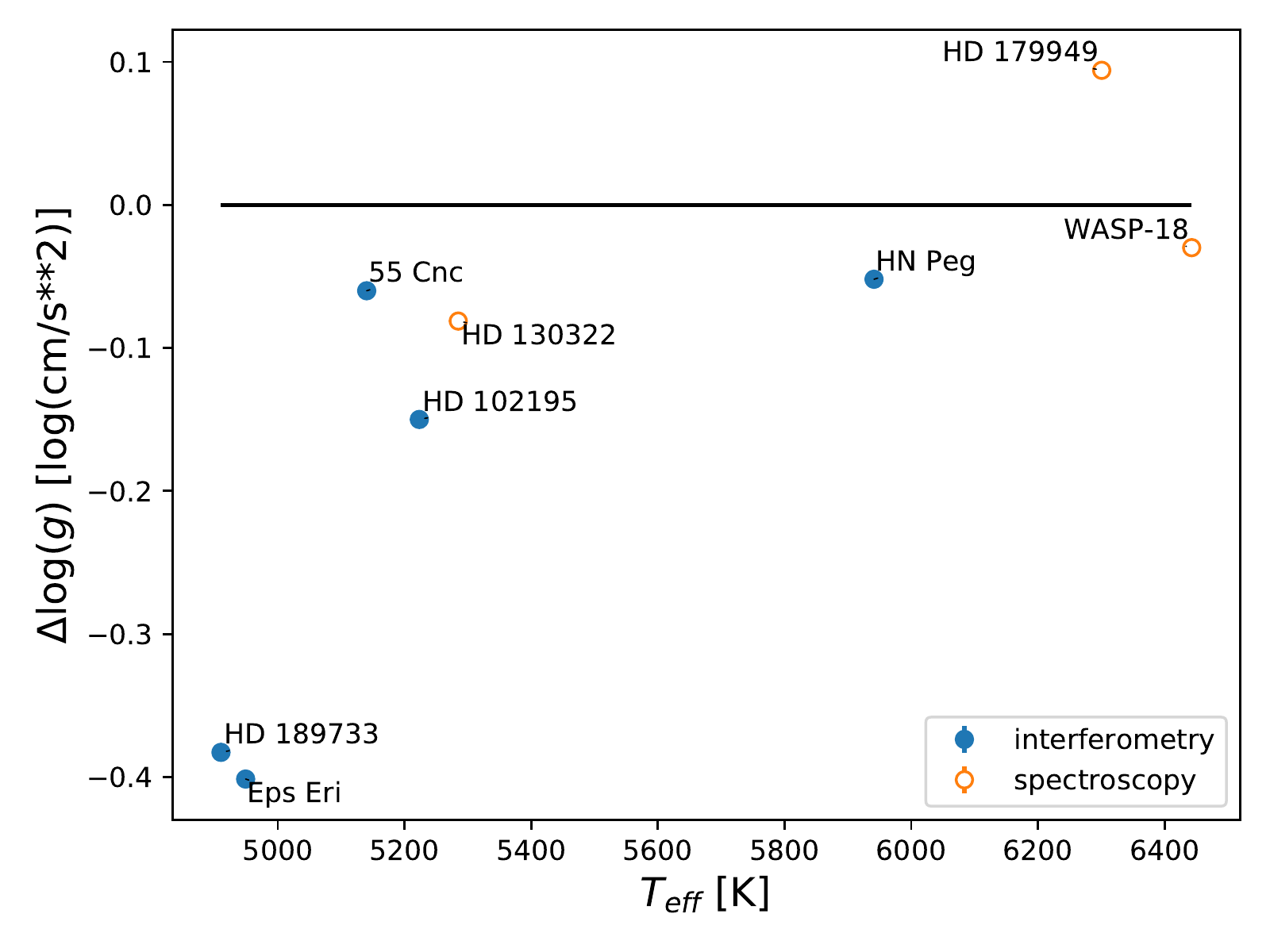}} \\
    \subfloat[\smonh]{\includegraphics[width=0.45\textwidth]{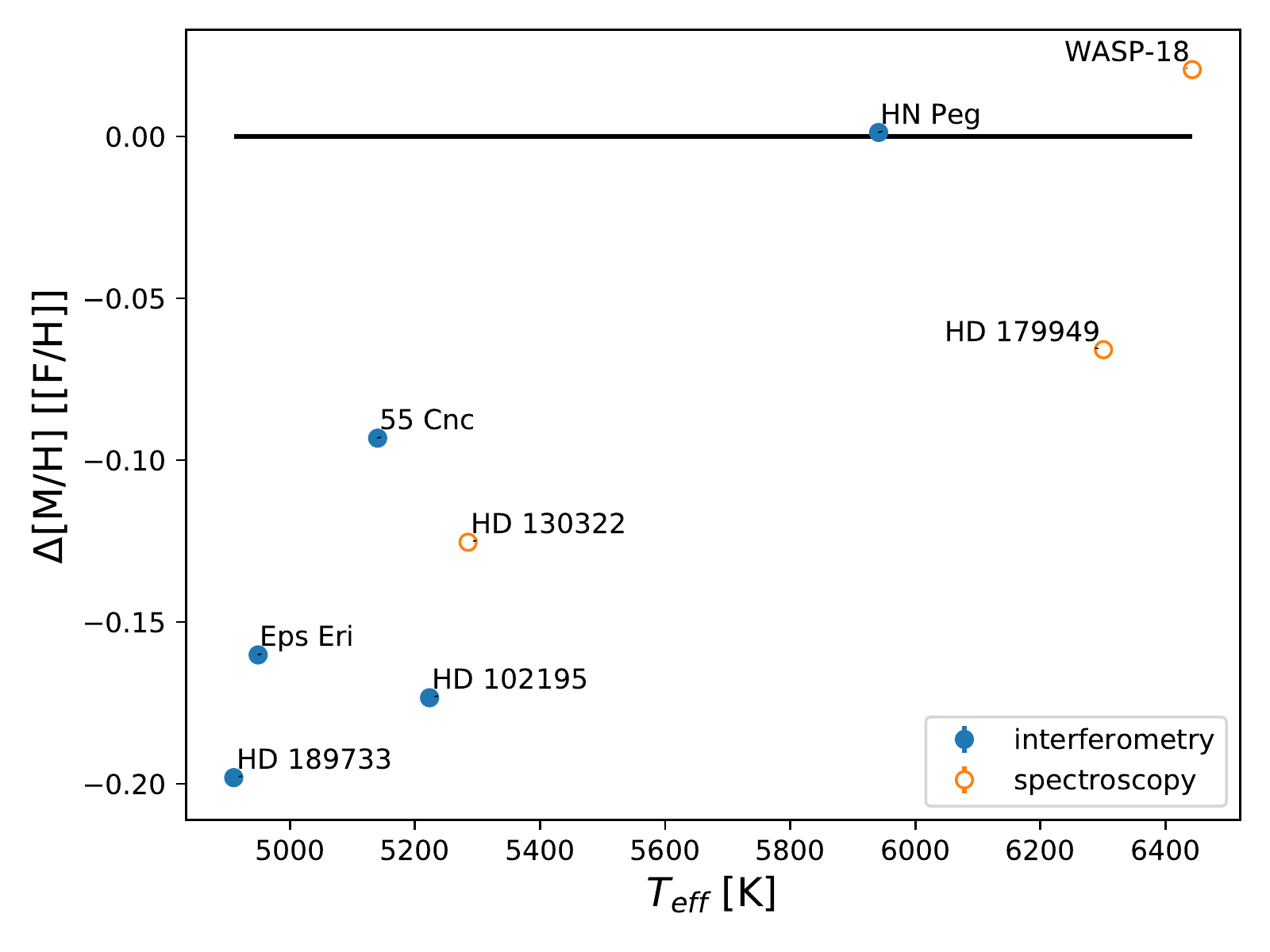}}
    \caption{Differences between the literature values and the parameter values derived in this paper. For stars represented by filled blue circles, \steff and \slogg literature values are derived from interferometry, while for stars with unfilled orange circles the literature values are derived from spectroscopy or photometry instead. Sources for all values are given in \autoref{tab:hd22049}.}
    \label{fig:delta}
\end{figure*}

It is useful to compare the stellar parameters we derived in this study with the parameters derived in other studies.
Where possible we used values derived from interferometric measurements, as those are independent from spectroscopic
or photometric methods. However that is only possible for \steff and \slogg, but not for metallicity. The numerical
values of our results and that of other studies are given in \autoref{tab:hd22049}, while we plot the differences in
\autoref{fig:delta}. 
We can see that our values agree mostly with the other studies, but there appears to be some dependence on \steff, especially for \steff\ itself. We also see a notable offset between our \smonh and \slogg values and those from other studies.

\section{Trends}
\label{sec:trends} 

\begin{figure*}[ht]
    \centering
    \subfloat[\steff]{\includegraphics[width=0.45\textwidth]{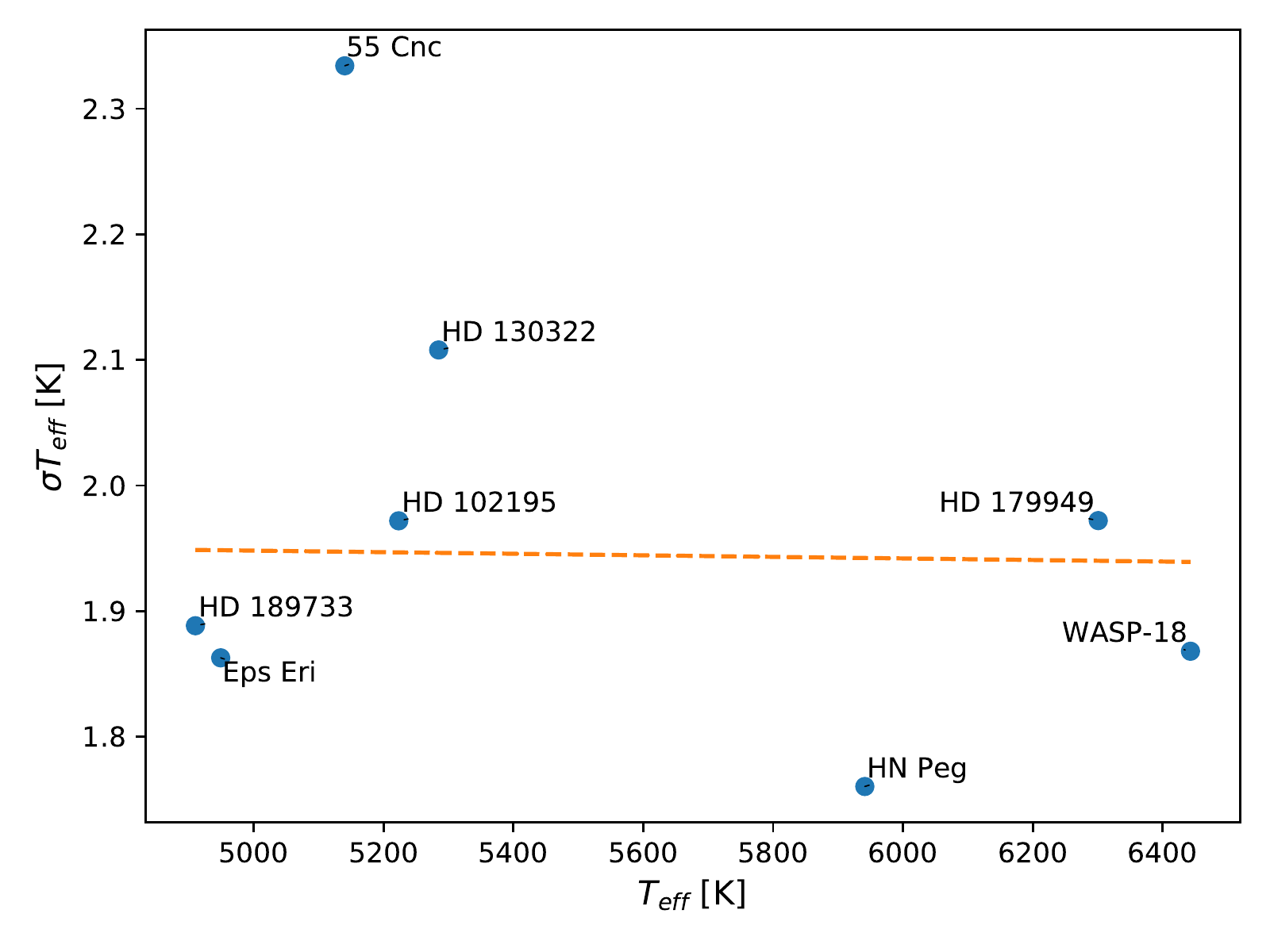}}
    \subfloat[\slogg]{\includegraphics[width=0.45\textwidth]{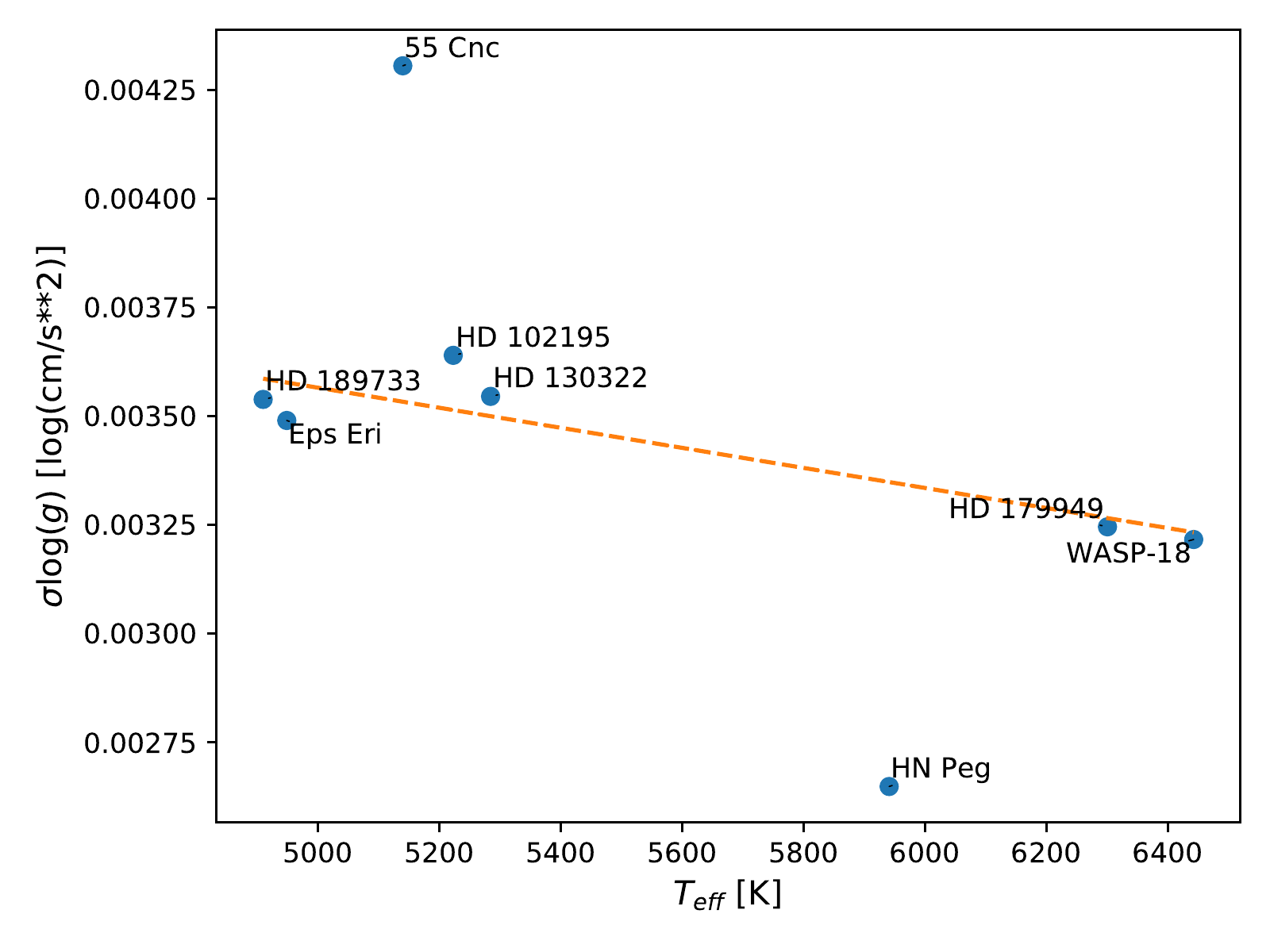}}\\
    \subfloat[\smonh]{\includegraphics[width=0.45\textwidth]{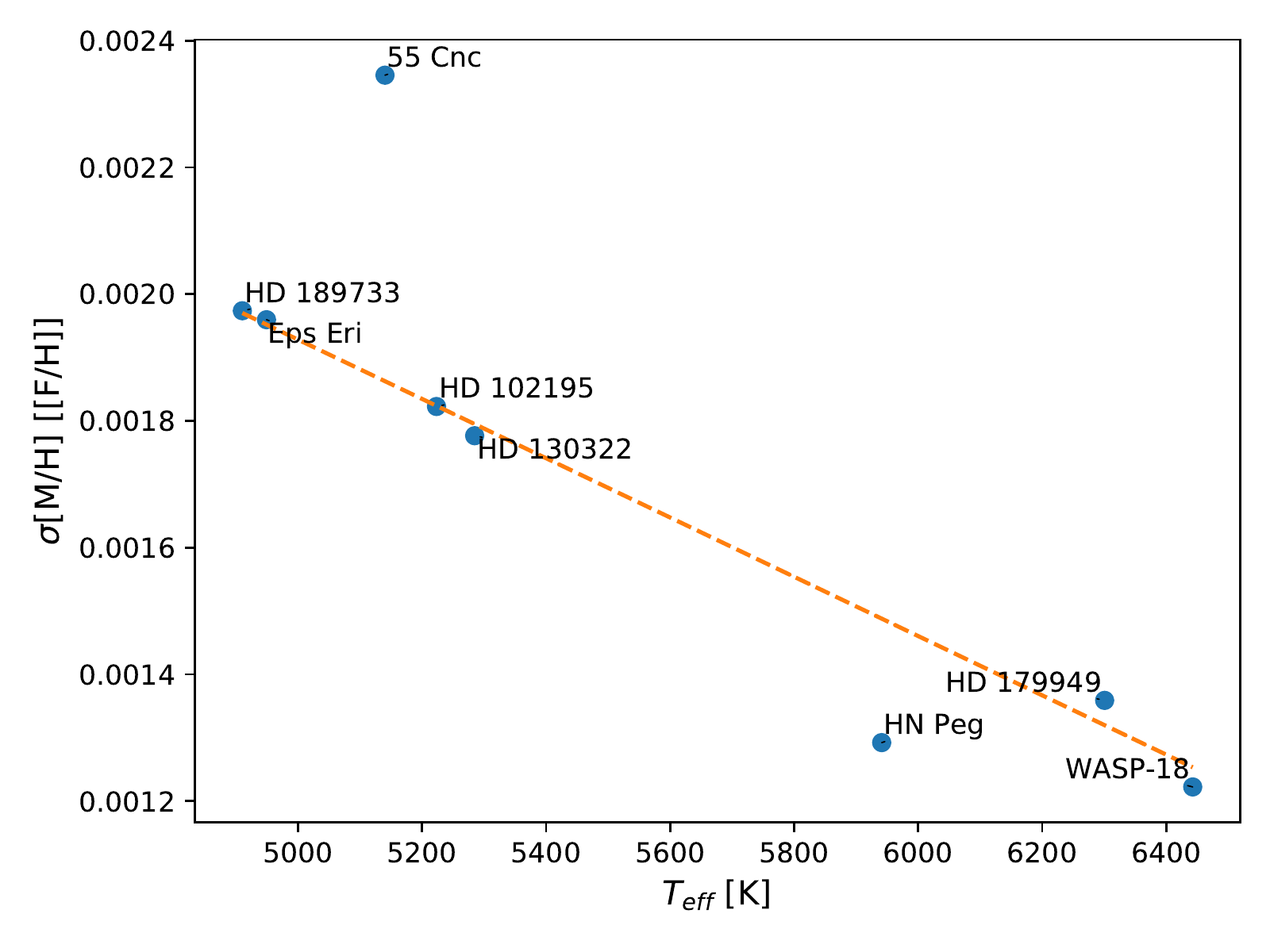}}
    \subfloat[\svmic]{\includegraphics[width=0.45\textwidth]{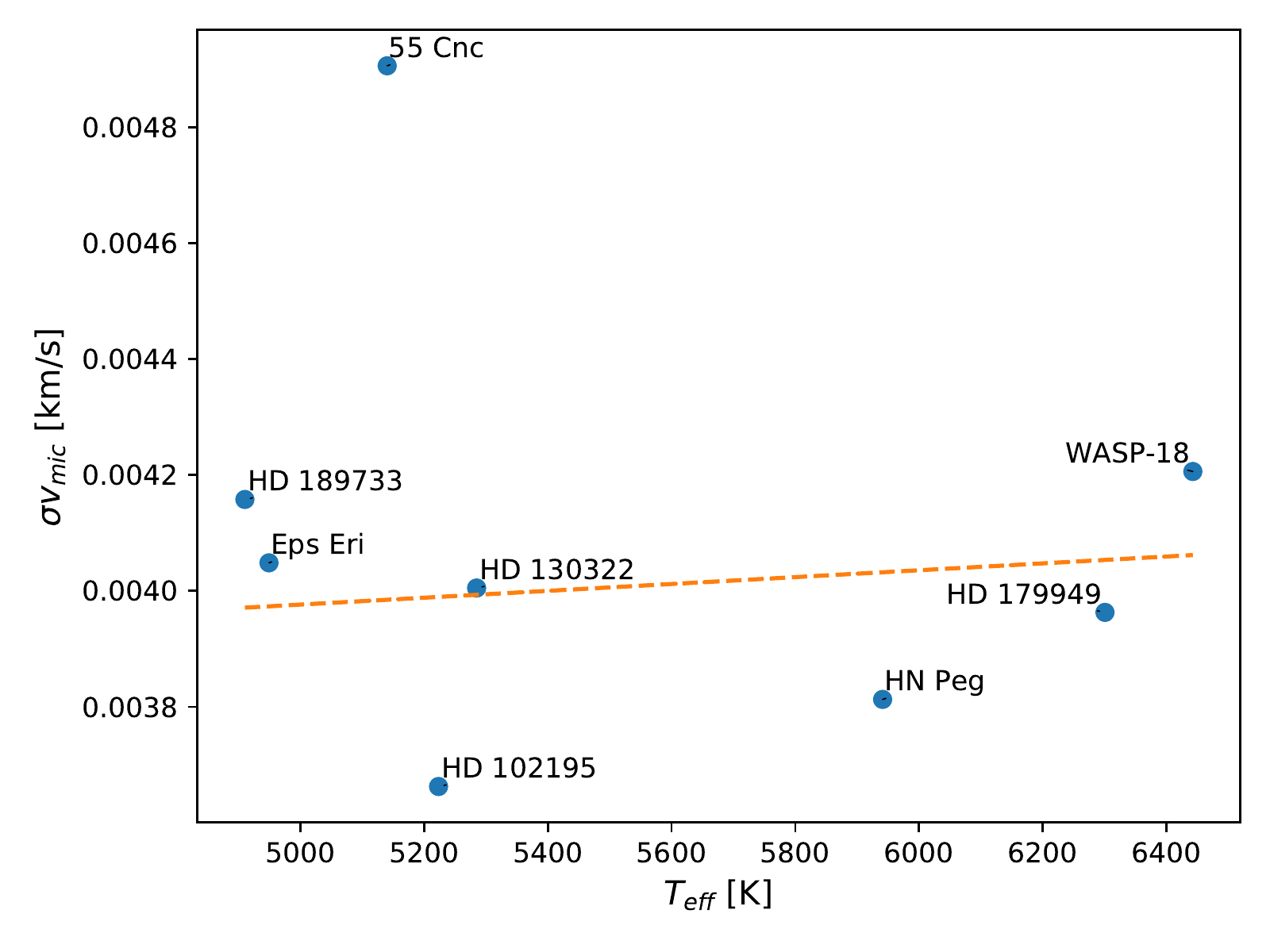}}\\
    \subfloat[\svmac]{\includegraphics[width=0.45\textwidth]{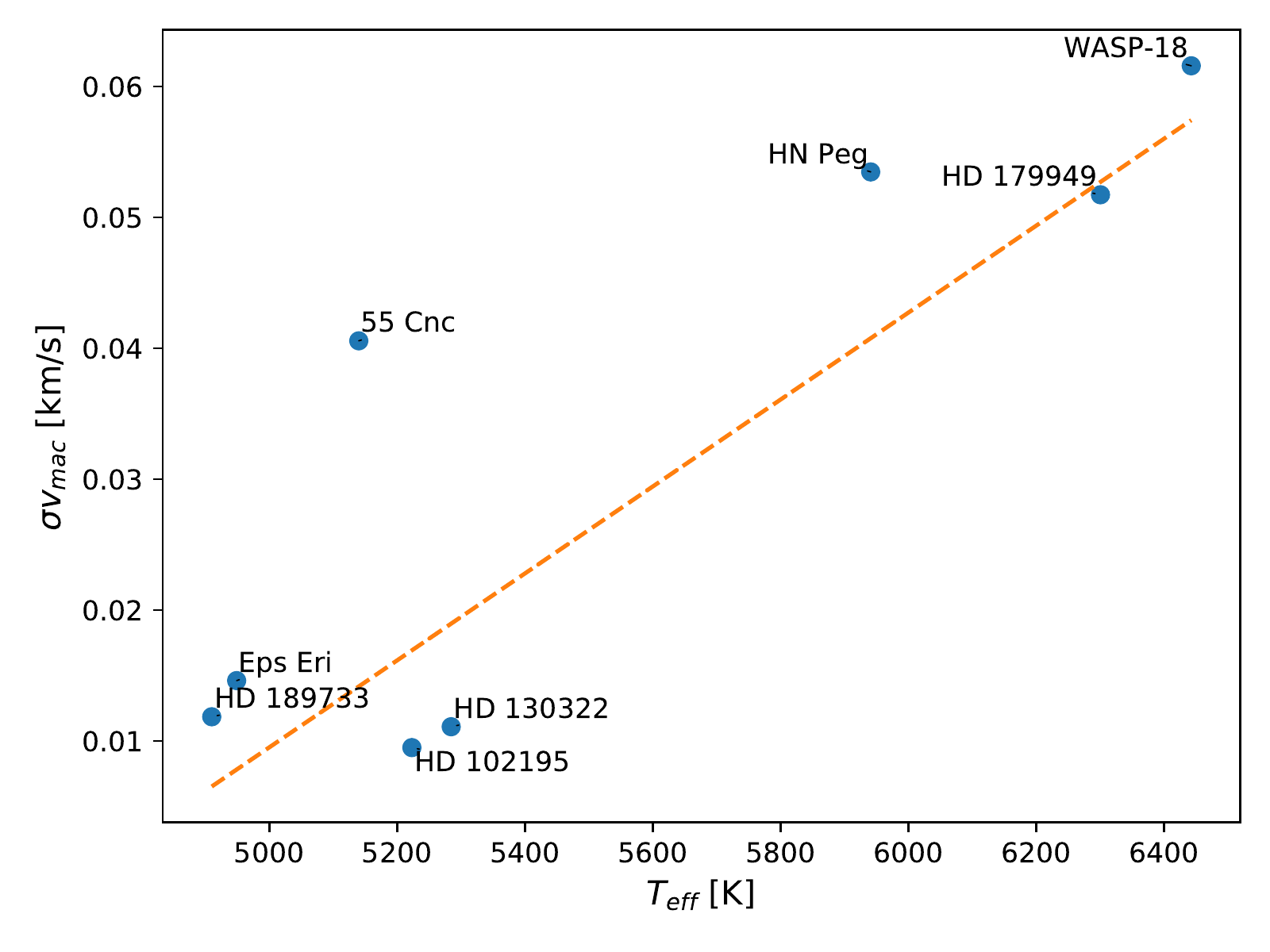}}
    \subfloat[\svsini]{\includegraphics[width=0.45\textwidth]{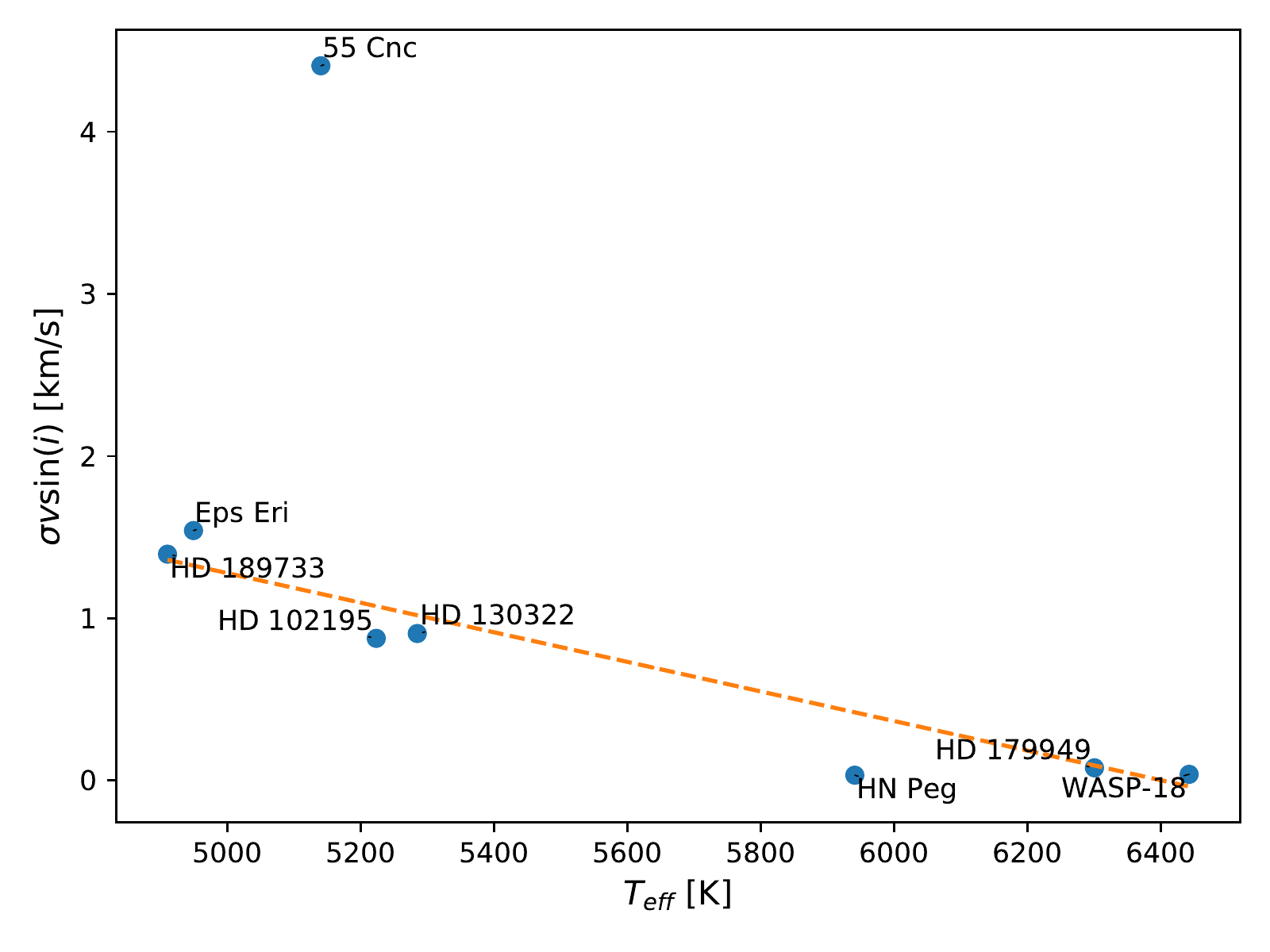}}
    \caption{Uncertainty trends with the stellar temperature. Each star is marked (blue solid circle) with their names. It also shows the best fit linear relationship (orange). \hnpeg and \cnc are ignored in the surface gravity fit.}
    \label{fig:trends}
\end{figure*}

Finally it is also interesting to investigate how the uncertainties of the different parameters depend on \steff of the star. We therefore plot this relationship in \autoref{fig:trends} for all stars. The uncertainty of the temperature depends only weakly on the temperature. Similarly the uncertainty of the surface gravity decreases slightly with temperature, while the uncertainty of the metallicity shows a stronger negative correlation. The uncertainty of the micro turbulence parameter shows no visible correlation with the temperature, while the uncertainty of the macro turbulence shows a clear upwards trend. The uncertainty of the rotation velocity decreases with increasing temperature. However there is a degeneracy with the resolution of the instrument, as below $\approx \SI{2}{\uvsini}$ the rotational broadening is smaller than the instrumental broadening. This increases the uncertainties of \svsini significantly for those cases.

\begin{figure*}[ht]
    \centering
    \subfloat[\svmic]{\includegraphics[width=0.45\textwidth]{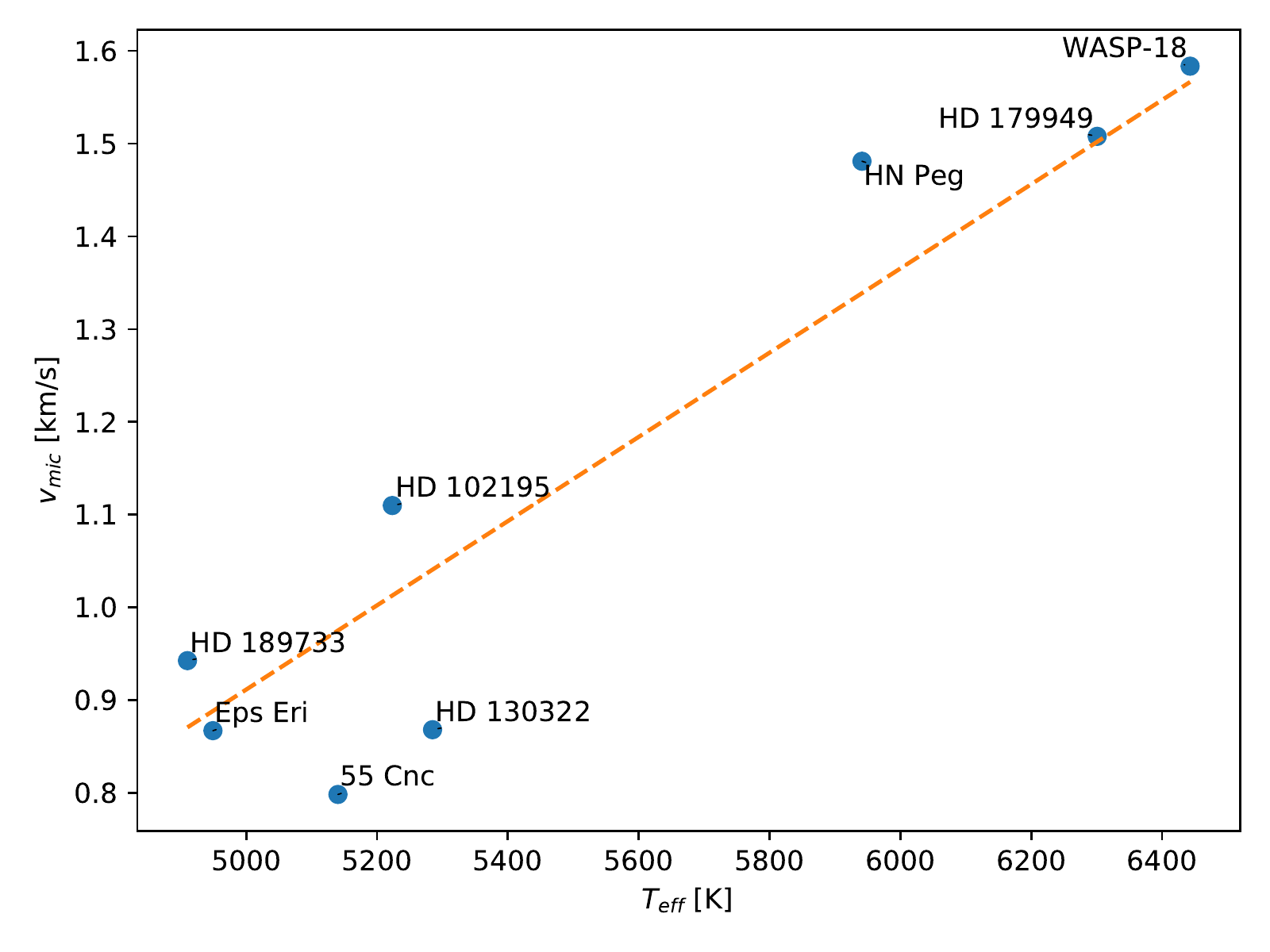}} 
    \subfloat[\svmac]{\includegraphics[width=0.45\textwidth]{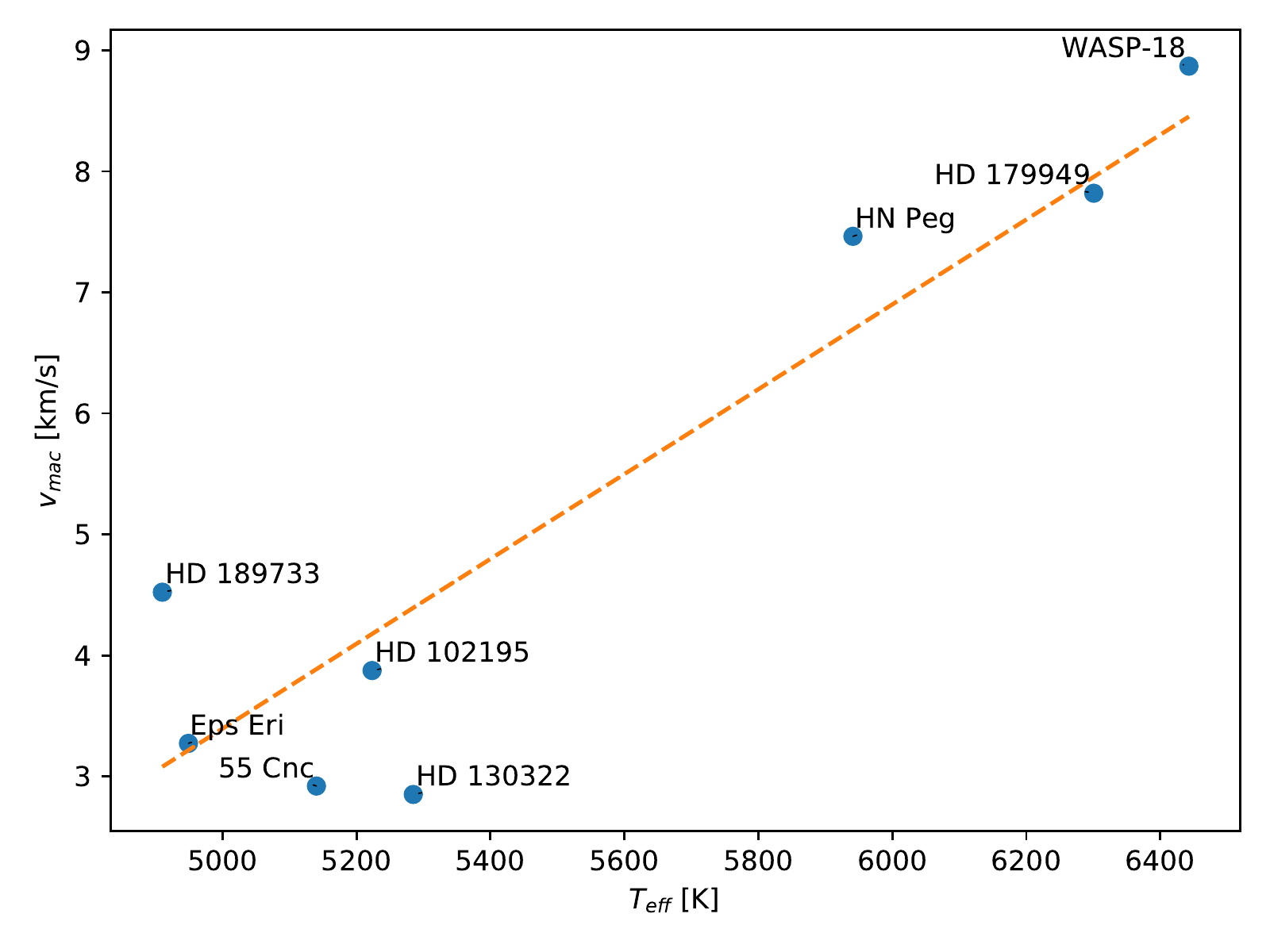}} \\
    \subfloat[\svsini]{\includegraphics[width=0.45\textwidth]{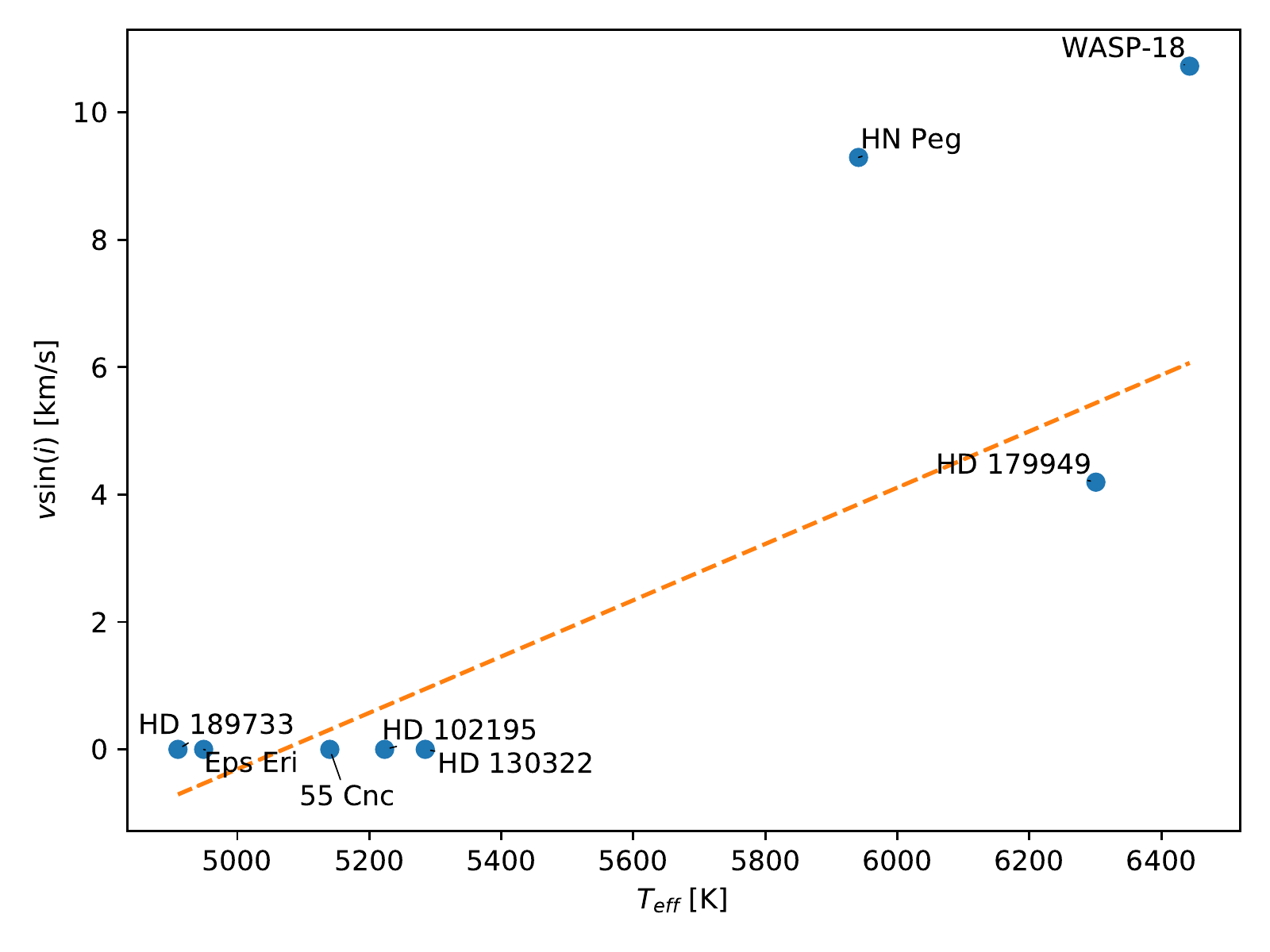}}
    \caption{Micro- and macroturbulence parameters, as well as the rotational velocity, of the star, as a function of the stellar temperature. The orange dashed line shows the linear regression fit to those values.}
    \label{fig:velocity_trends}
\end{figure*}

Additionally we also compare the values of the turbulence parameters and the stellar rotation as a function of the stellar temperature in \autoref{fig:velocity_trends}. As expected the micro- and macroturbulence increase with the stellar temperature. The rotational velocity also increases with the stellar temperature, since rotational velocity decreases along the main sequence towards earlier spectral types \citep{2005oasp.book.....G}.

\section{Conclusion}
\label{sec:conclusion}
We have shown that \pysme is a suitable successor for \idlsme. We have additionally derived stellar parameters with \pysme for a number of exoplanet host stars and see that their values agree both with values derived from spectroscopy, as well as independent interferometric values from other studies.

\begin{acknowledgements}
This research has made use of the services of the ESO Science Archive Facility. 
Based on observations collected at the European Southern Observatory under ESO programs 60.A-9036(A), 192.C-0224(A), 083.C-0794(A), 072.C-0488(E), 288.C-5010(A), 0104.C-0849(A), and 098.C-0739(A). 
This work has made use of the VALD database, operated at Uppsala University, the Institute of Astronomy RAS in Moscow, and the University of Vienna. \\
\textit{Software:} PySME relies on a number of Python packages to function , these are NumPy \citep{harris2020array}, SciPy \citep{2020SciPy-NMeth}, Astropy \citep{astropy:2013,astropy:2018}, Matplotlib \citep{4160265}, and pandas \citep{mckinney-proc-scipy-2010}
\end{acknowledgements}

\clearpage
\begin{landscape}
\begin{table}
\raggedbottom
\begin{tabular}[t]{lccc}
    \toprule
    Parameter [Unit] & {NEXA} & {Interferometry} & {\pysme} \\[0.5ex]
    \midrule
    \epseri & & & \\[0.5ex]
    \midrule
    \steff~[\uteff] & \val{5065}[131][72] [1] & \val{5077}[35] [2] & $\val{4949}[2] \pm 191$ \\[0.5ex]
    \slogg~[\ulogg] & \val{4.61}[0.06][0.09] [1] & \val{4.61}[0.02] [2] & $\val{4.21}[0.003] \pm 0.63$ \\[0.5ex]
    \smonh~[\umonh] & \val{-0.05}[0.10] [1] & \val{-0.06}[0.1] [2] & $\val{-0.22}[0.002] \pm 0.25$ \\[0.5ex]
    \svmic~[\uvmic] & (\val{1}) & - & $\val{0.87}[0.004] \pm 0.7$ \\[0.5ex]
    \svmac~[\uvmac] & (\val{4}) & - & $\val{3.27}[0.01] \pm 1.8$  \\[0.5ex]
    \svsini~[\uvsini] & \val{2.4}[0.5] [3] & - & $\val{0.0}[4.4] \pm 189$ \\[0.5ex]
    \midrule
    \hnpeg & & & \\[0.5ex]
    \midrule
    \steff~[\uteff] & \val{5927}[136][155] [1] & \val{5860}[83] [2] & $\val{5941}[2] \pm 222$ \\[0.5ex]
    \slogg~[\ulogg] & \val{4.46}[0.08][0.09] [1] & \val{4.43}[0.05] [2] & $\val{4.37}[0.003] \pm 0.61$ \\[0.5ex]
    \smonh~[\umonh] & \val{-0.03}[0.01] [1] & \val{-0.16}[0.1] [2] & $\val{-0.16}[0.001] \pm 0.18$ \\[0.5ex]
    \svmic~[\uvmic] & (\val{1}) & - & $\val{1.48}[0.004] \pm 0.6 $ \\[0.5ex]
    \svmac~[\uvmac] & (\val{4}) & - & $\val{7.46}[0.05] \pm 5.2 $\\[0.5ex]
    \svsini~[\uvsini] & \val{8.73}[0.06][0.05] [4] & - & $\val{9.29}[0.03] \pm 3.4$\\[0.5ex]
    \midrule
    \hdten & & & \\[0.5ex]
    \midrule
    \steff~[\uteff] & \val{5276}[90][110] [1] & \val{5277}[60] [2] &  $\val{5223}[2] \pm 197$ \\[0.5ex]
    \slogg~[\ulogg] & \val{4.55}[0.08][0.07] [1] & \val{4.50}[0.05] [2] & $\val{4.35}[0.004] \pm 0.64$\\[0.5ex]
    \smonh~[\umonh] & \val{0.05}[0.1] [1] & \val{0.1}[0.05] [2] & $\val{-0.07}[0.002] \pm 0.23$ \\[0.5ex]
    \svmic~[\uvmic] & (\val{1}) & - & $\val{1.11}[0.004] \pm 0.6$ \\[0.5ex]
    \svmac~[\uvmac] & (\val{4}) & - & $\val{3.87}[0.01] \pm 1.8$ \\[0.5ex]
    \svsini~[\uvsini] & \val{2.6} [5] & - & $\val{0.0}[0.9] \pm 220$ \\[0.5ex]
    \midrule
    \hdthirteen & & & \\
    \midrule
    \steff~[\uteff] & \val{5377}[132][87] [1] & - & $\val{5285}[2] \pm 212$ \\[0.5ex]
    \slogg~[\ulogg] & \val{4.52}[0.06][0.09] [1] & - & $\val{4.44}[0.004] \pm 0.60$ \\[0.5ex]
    \smonh~[\umonh] & \val{0.02}[0.1] [1] & - & $\val{-0.11}[0.002] \pm 0.22$ \\[0.5ex]
    \svmic~[\uvmic] & (\val{1}) & - & $\val{0.87}[0.004] \pm 0.6$ \\[0.5ex]
    \svmac~[\uvmac] & (\val{4}) & - & $\val{2.85}[0.01] \pm 1.8$ \\[0.5ex]
    \svsini~[\uvsini] & \val{0.5}[0.5] [6] & - & $\val{0.0}[0.9] \pm 167$ \\[0.5ex]
    \bottomrule

    \end{tabular}
\quad
    \begin{tabular}[t]{lccc}
     \toprule
    Parameter [Unit] & {NEXA} & {Interferometry} & {\pysme} \\[0.5ex]
    \midrule
    \hdseventeen & & & \\[0.5ex]
    \midrule
    \steff~[\uteff] & \val{6148}[165][128] [1] & - & $\val{6301}[2] \pm 242$ \\[0.5ex]
    \slogg~[\ulogg] & \val{4.32}[0.08][0.10] [1] & - & $\val{4.41}[0.003] \pm 0.80$ \\[0.5ex]
    \smonh~[\umonh] & \val{0.2}[0.1] [1] & - & $\val{0.13}[0.001] \pm 0.19$ \\[0.5ex]
    \svmic~[\uvmic] & (\val{1}) & - & $\val{1.51}[0.004] \pm 0.5$\\[0.5ex]
    \svmac~[\uvmac] & (\val{4}) & - & $\val{7.82}[0.05] \pm 3.0$\\[0.5ex]
    \svsini~[\uvsini] & \val{6.52} [7] & - & $\val{4.19}[0.07] \pm 4.7$ \\[0.5ex]
    \midrule
    \hdeighteen & & & \\[0.5ex]
    \midrule
    \steff~[\uteff] & \val{5023}[165][119] [1] & \val{5024}[60] [2] & $\val{4910}[2] \pm 203$ \\[0.5ex]
    \slogg~[\ulogg] & \val{4.58}[0.08][0.12] [1] & \val{4.51}[0.05] [2] & $\val{4.12}[0.004] \pm 0.65$\\[0.5ex]
    \smonh~[\umonh] & \val{0.20}[0.10] [1] & \val{0.07}[0.05] [2] & $\val{-0.13}[0.002] \pm 0.27$\\[0.5ex]
    \svmic~[\uvmic] & (\val{1}) & - & $\val{0.94}[0.004] \pm 0.7$\\[0.5ex]
    \svmac~[\uvmac] & (\val{4}) & - & $\val{4.52}[0.01] \pm 2.1$\\[0.5ex]
    \svsini~[\uvsini] & \val{3.5}[1] [8] & - & $\val{0.0}[1.4] \pm 289$ \\[0.5ex]
    \midrule
    \cnc & & & \\[0.5ex]
    \midrule
    \steff~[\uteff] & \val{5250}[123][172] [1] & \val{5172}[18] [2] & $\val{5140}[2] \pm 228$ \\[0.5ex]
    \slogg~[\ulogg] & \val{4.42}[0.05][0.14] [1] & \val{4.43}[0.02] [2] & $\val{4.37}[0.004] \pm 0.74$ \\[0.5ex]
    \smonh~[\umonh] & \val{0.35}[0.10] [1] & \val{0.35}[0.10] [2] & $\val{0.26}[0.002] \pm 0.28$ \\[0.5ex]
    \svmic~[\uvmic] & (\val{1}) & - & $\val{0.80}[0.005] \pm 0.7$ \\[0.5ex]
    \svmac~[\uvmac] & (\val{4}) & - & $\val{2.92}[0.04] \pm 1.8$ \\[0.5ex]
    \svsini~[\uvsini] & < \val{1.23}[0.01] [9] & - & $\val{0.0}[4.4] \pm 189$ \\[0.5ex]
    \midrule
    \pagebreak
    \wasp & & & \\[0ex]
    \midrule
    \steff~[\uteff] & \val{6226}[162][117] [1] & - & $\val{6443}[2] \pm 328$\\[0.5ex]
    \slogg~[\ulogg] & \val{4.26}[0.10][0.07] [1] & - & $\val{4.23}[0.003] \pm 1.01$ \\[0.5ex]
    \smonh~[\umonh] & \val{0.1}[0.1] [1] & - & $\val{0.12}[0.001] \pm 0.22$ \\[0.5ex]
    \svmic~[\uvmic] & (\val{1}) & - & $\val{1.58}[0.004] \pm 0.7$ \\[0.5ex]
    \svmac~[\uvmac] & (\val{4}) & - & $\val{8.87}[0.06] \pm 7.3$ \\[0.5ex]
    \svsini~[\uvsini] & \val{11}[1.5] [8] & - & $\val{10.7}[0.04] \pm 4.7$ \\[0.5ex]
    %
    \bottomrule
    \end{tabular}
    
    \caption{Stellar parameters of all stars investigated in this paper. For \pysme both the fit uncertainty and the model uncertainty are given (see \autoref{sec:uncertainties}). Two sets of literature values are given. The first is from the NASA Exoplanet Archive (NEXA). Those are used as initial values for the fitting procedure. The micro- and macroturbulence parameters are given in brackets () as there are no existing literature values, instead we use general estimates. In the second set of values the \steff and \slogg are derived by interferometry if that data is available, or spectroscopy otherwise. These are used in the comparison in \autoref{sec:trends}. Note that for \cnc the analysis is independent of the \idlsme analysis discussed in \autoref{sec:targets}, which used a different continuum normalization, linelist, and NLTE coefficients, but instead follows the same steps as for the other targets. References: [1] {\citet{2019AJ....158..138S}}, [2] {\citet{2017ApJ...836...77Y}}, [3] {\citet{2005ApJS..159..141V}}, [4] {\citet{2021AJ....161..114S}}, [5] {\citet{2007A&A...467..721M}}, [6] {\citet{2015ApJ...803....8H}}, [7] {\citet{2014A&A...570A..80T}}, [8] {\citet{2017A&A...602A.107B}}, [9] {\citet{2018A&A...619A...1B}}, [10] {\citet{2018A&A...616A...1G}}, [11] {\citet{2021A&A...649A.177M}}, [12] {\citet{2020Natur.582..497P}} }
    \label{tab:hd22049}
\end{table}
\end{landscape}

\clearpage
\bibliography{main.bib}

\clearpage
\appendix

\section{Data Access}
\label{sec:data_access}
The input spectra are available from the ESO Archive Science Portal \url{http://archive.eso.org/scienceportal/} with the program IDs and archive IDs given in \autoref{tab:datasets}.
All generated spectra are available as SME files from Zenodo servers using \href{https://doi.org/10.5281/zenodo.6701350}{DOI 10.5281/zenodo.6701350}

\section{Installation}
\label{sec:installation}
PySME has been developed with usability in mind. We therefore decided to make the installation process as convenient as possible. For this purpose we provide compiled versions of the SME libraries for the most common environments (windows, mac osx and linux (using the manylinux2010 specification)). Additionally we have prepared the installation via PyPi so that PySME can be installed using pip using the following command:
\begin{verbatim}
    pip install pysme-astro
\end{verbatim}
The distribution on PyPi is automatically updated with any changes made to the open source Github repository.

In case the pre-compiled libraries are not compatible with your system it is also easy to compile the SME library from the source code.

\begin{verbatim}
git clone https://github.com/AWehrhahn/SMElib.git
cd SMElib
./bootstrap
./configure --prefix=$PWD
make install
\end{verbatim}

This will download and compile the code on your system. Afterwards you should find the library in the \opt{lib} (or \opt{bin} for windows) directory. Simply copy it into your installation of PySME.

\section{Uncertainties}
\label{sec:uncertainties_appendix}
Probability distributions for the stellar parameters derived for \epseri, as discussed in \autoref{sec:uncertainties}.

\begin{figure*}[ht]
    \centering
    \subfloat[\steff]{\includegraphics[width=0.9\columnwidth]{img/cumulative_probability_teff.pdf}}
    \subfloat[\steff]{\includegraphics[width=0.9\columnwidth]{img/probability_density_teff.pdf}}\\
    \subfloat[\slogg]{\includegraphics[width=0.9\columnwidth]{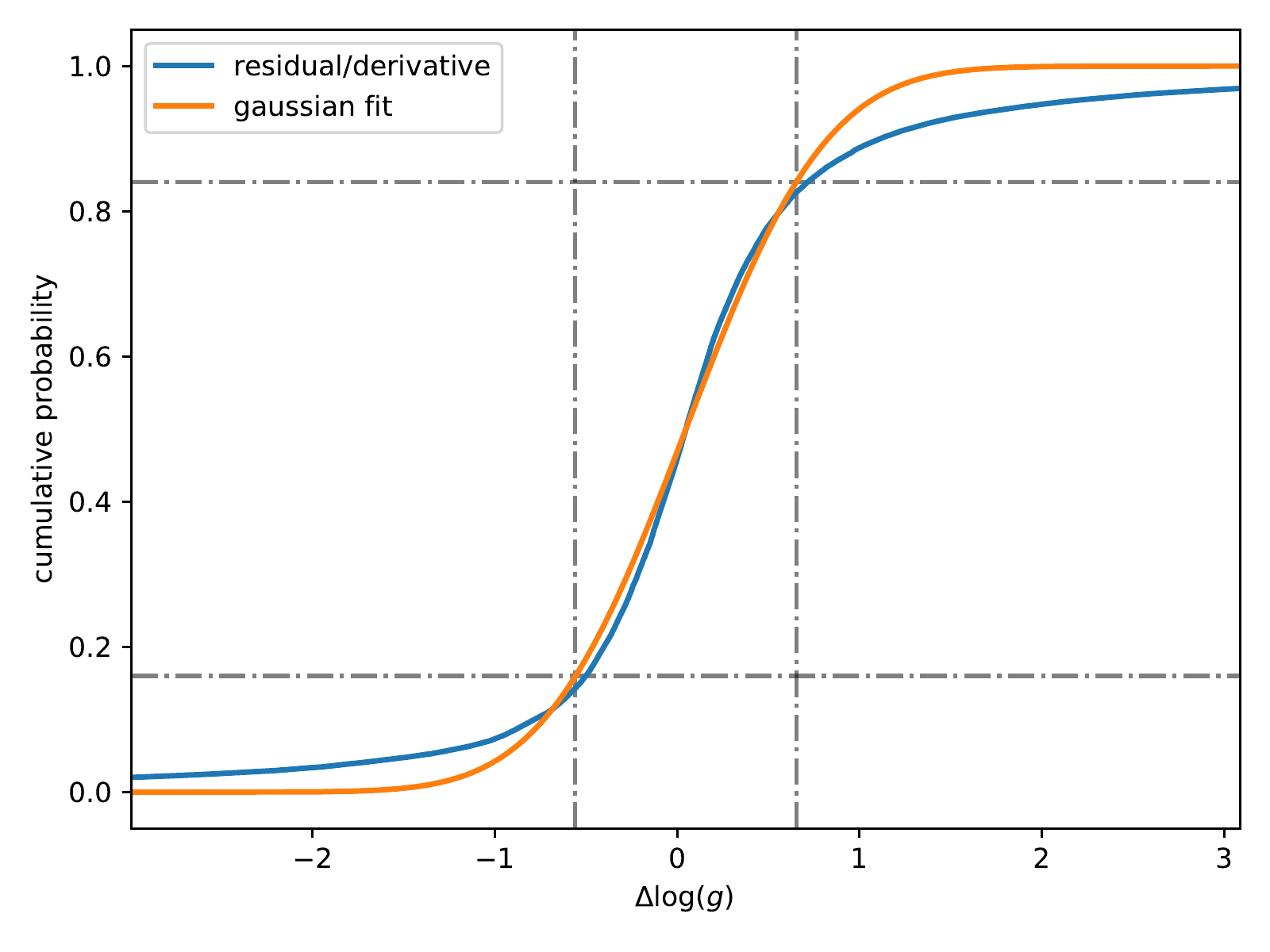}}
    \subfloat[\slogg]{\includegraphics[width=0.9\columnwidth]{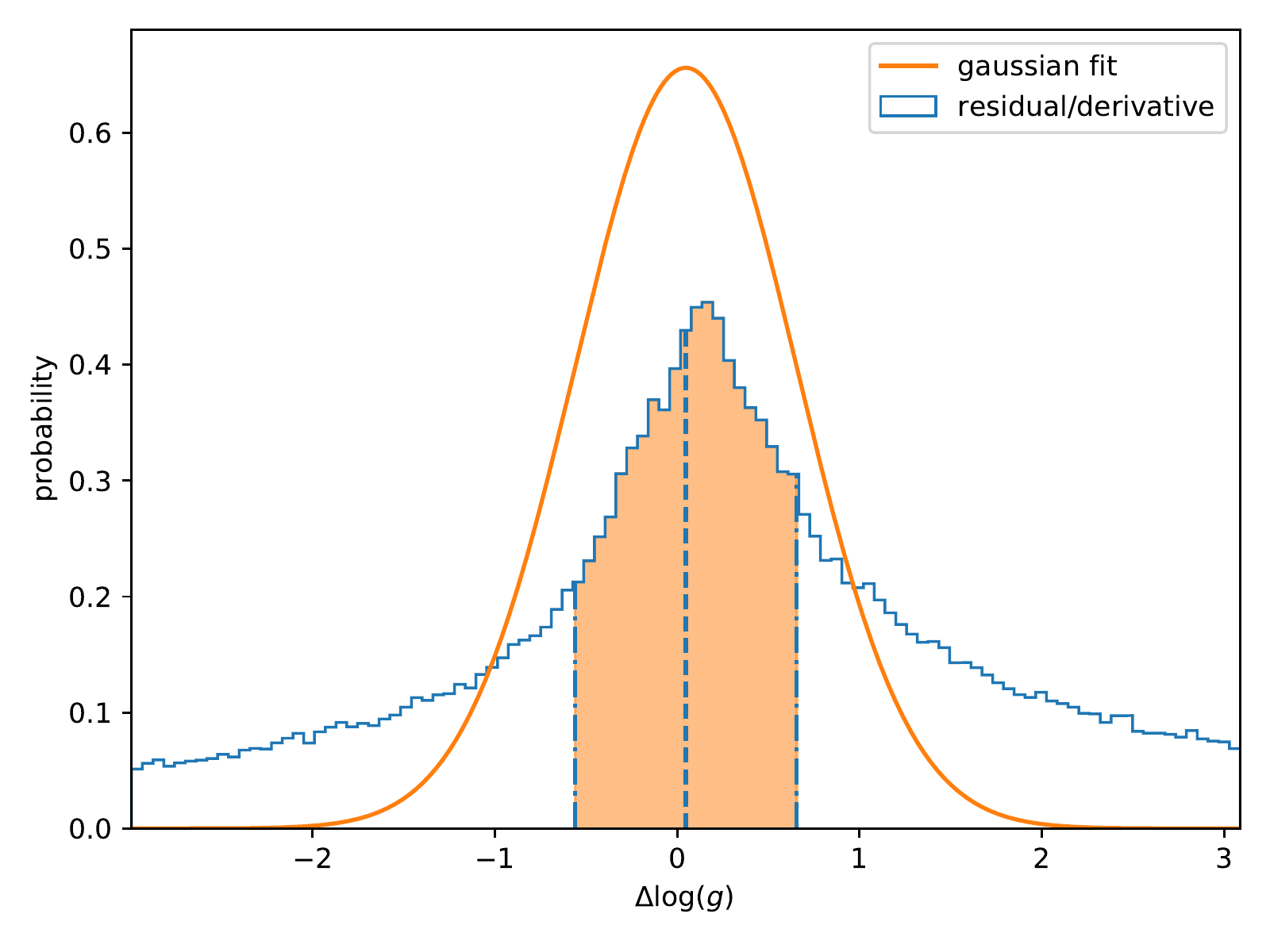}}\\
    \subfloat[\smonh]{\includegraphics[width=0.9\columnwidth]{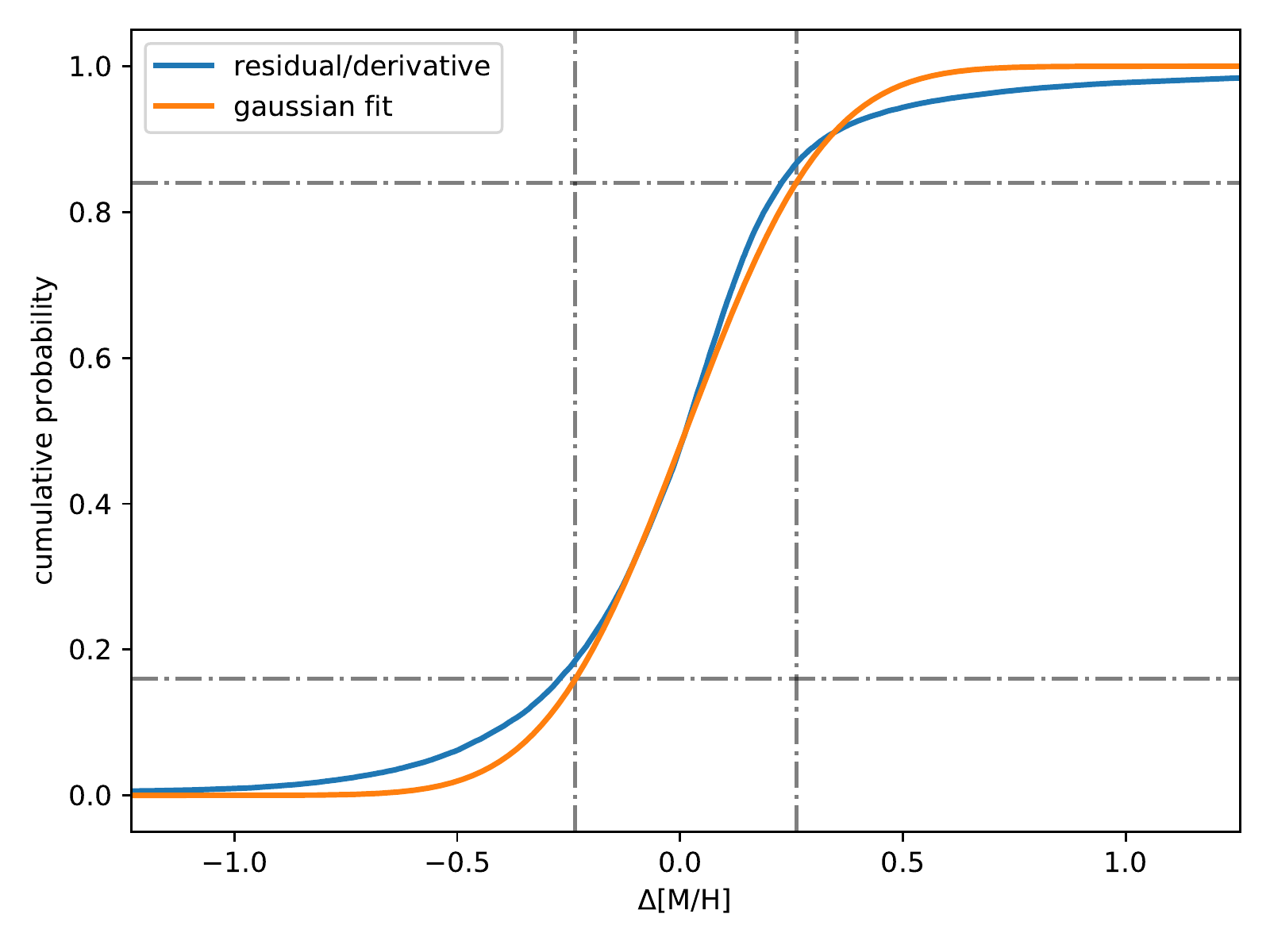}}
    \subfloat[\smonh]{\includegraphics[width=0.9\columnwidth]{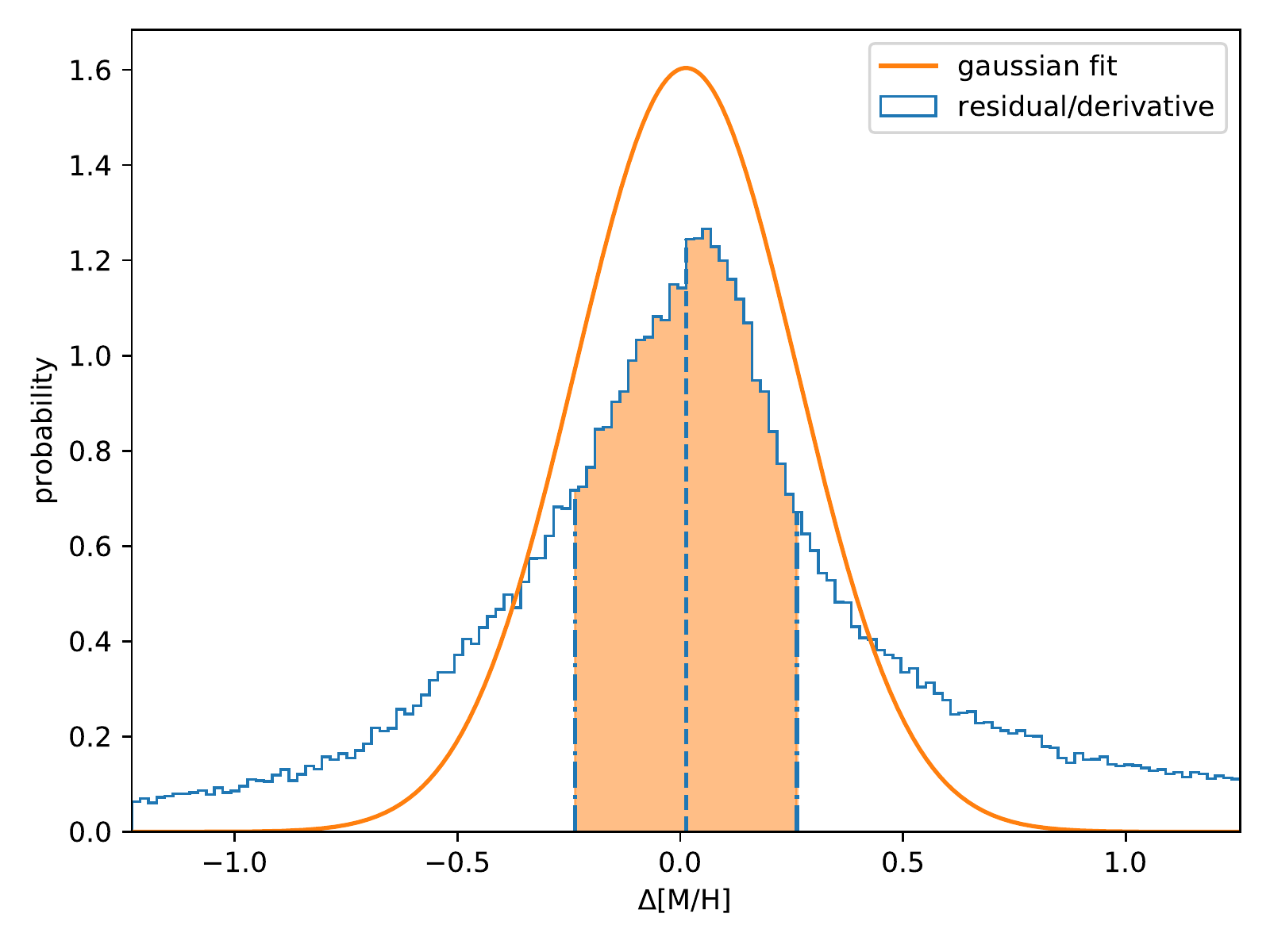}}
    \caption{Cumulative probability (left) and probability density (right) distribution of the fitting parameters for the \epseri spectrum derived from the fit results (blue). For comparison we also show the distribution of a Gaussian with the same median and $\sigma$ as derived from the cumulative probability. The median is marked by a dashed line (in density plots only), and the $1\sigma$~range is shown as a shaded region and dash dotted lines.}
    \label{fig:cum_prob_logg_app}
\end{figure*}

\begin{figure*}[ht]
    \centering
    \subfloat[\svmic]{\includegraphics[width=0.9\columnwidth]{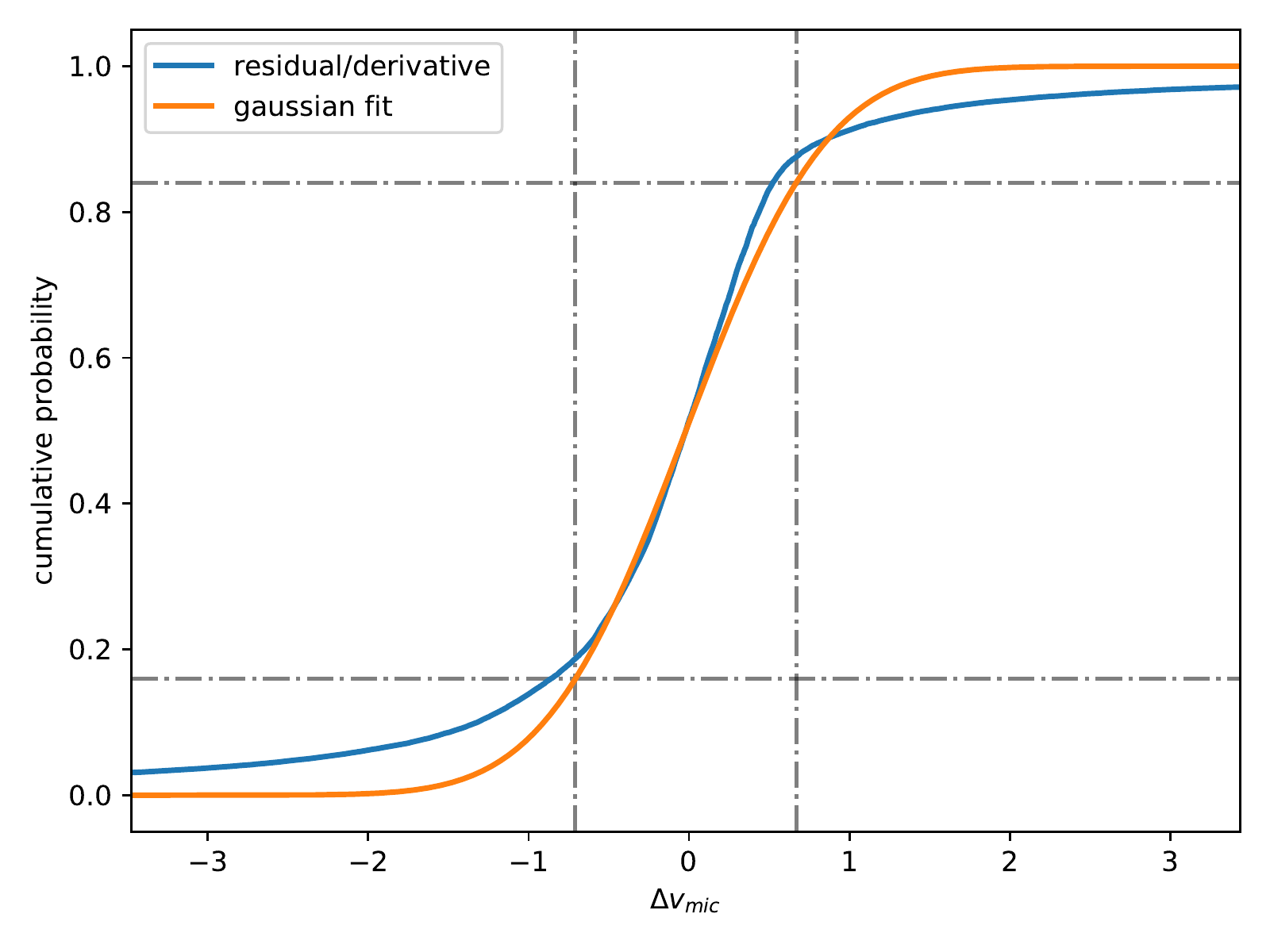}}
    \subfloat[\svmic]{\includegraphics[width=0.9\columnwidth]{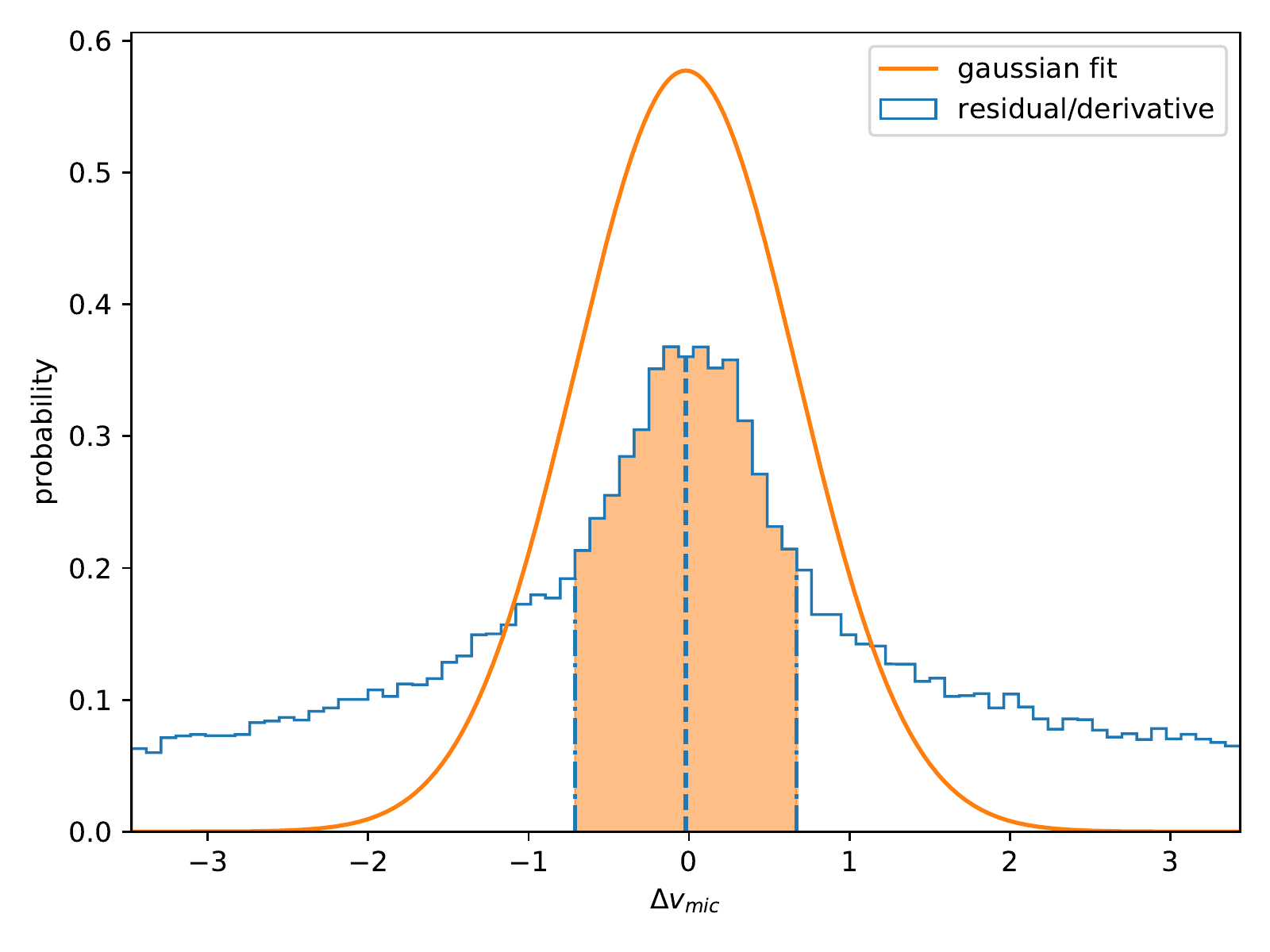}}\\
    \subfloat[\svmac]{\includegraphics[width=0.9\columnwidth]{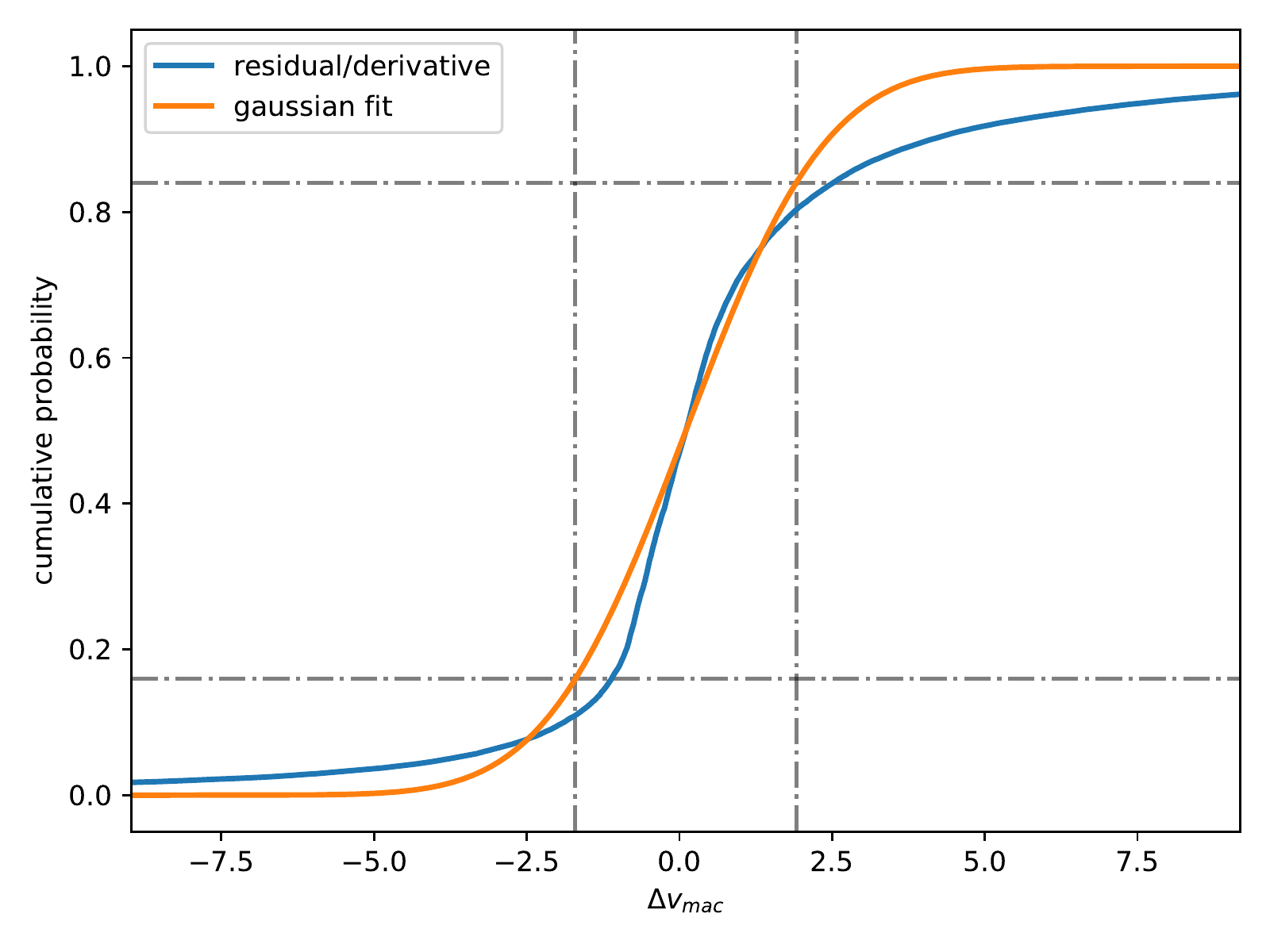}}
    \subfloat[\svmac]{\includegraphics[width=0.9\columnwidth]{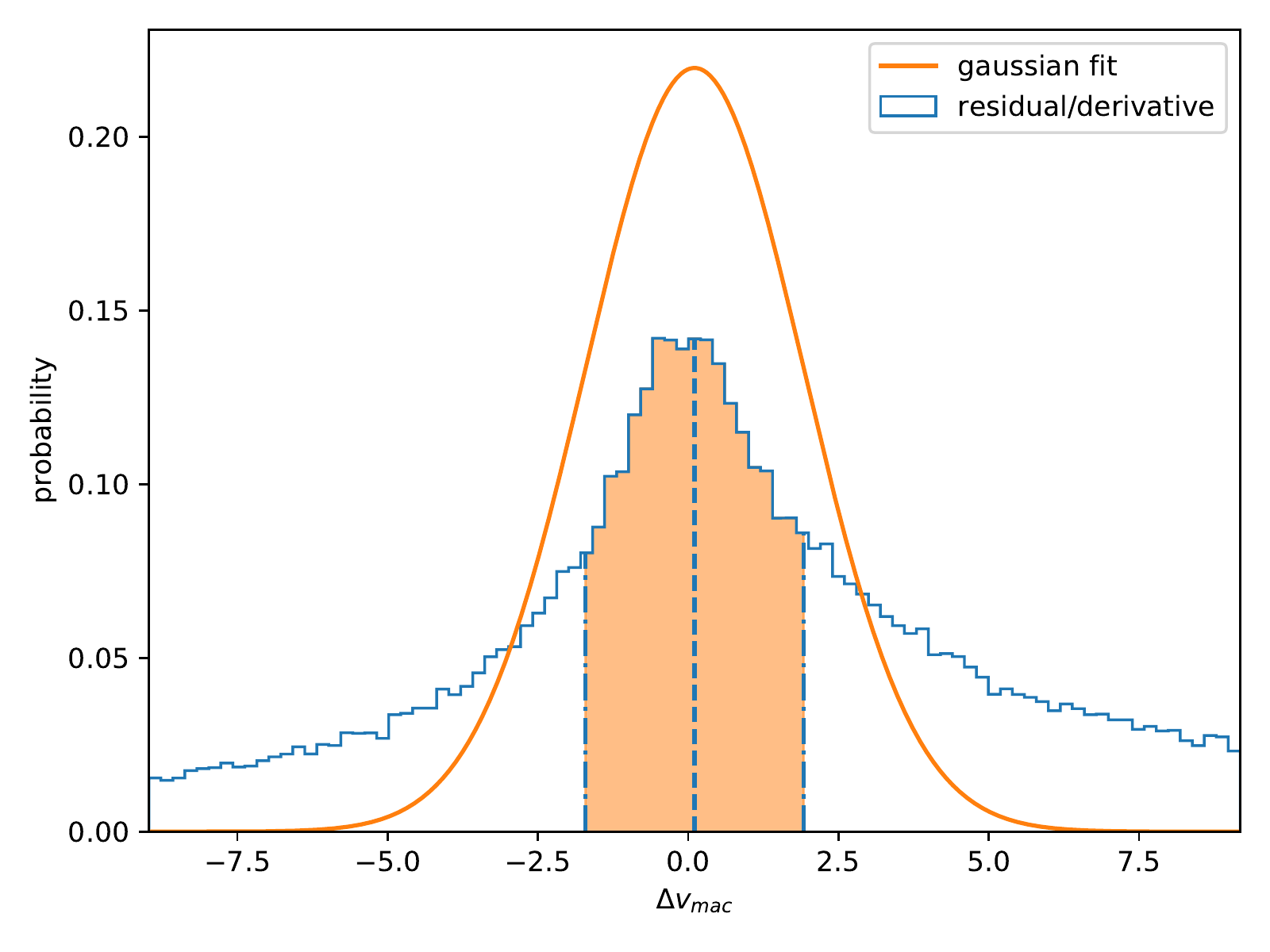}}\\
    \subfloat[\svsini]{\includegraphics[width=0.9\columnwidth]{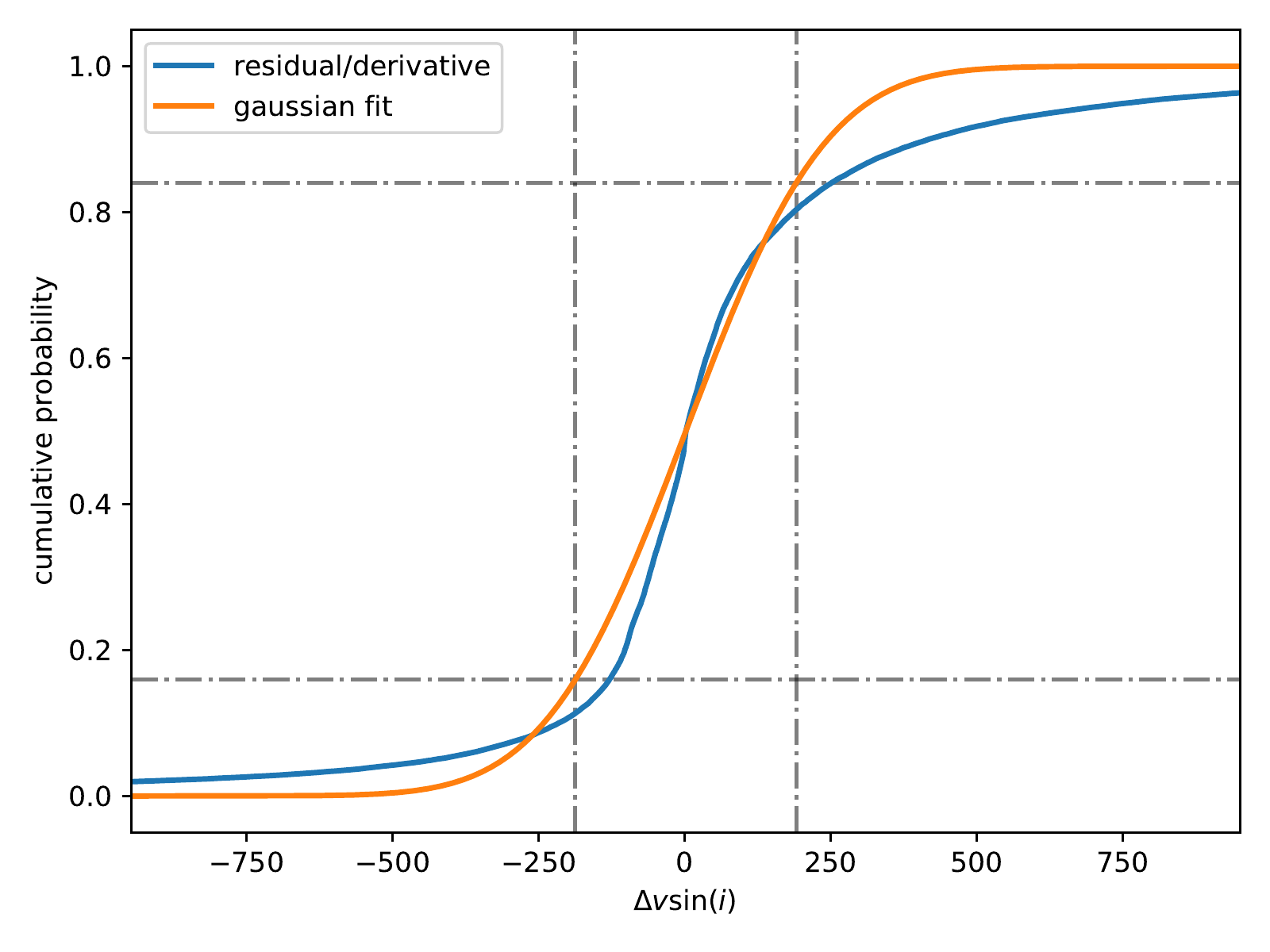}}
    \subfloat[\svsini]{\includegraphics[width=0.9\columnwidth]{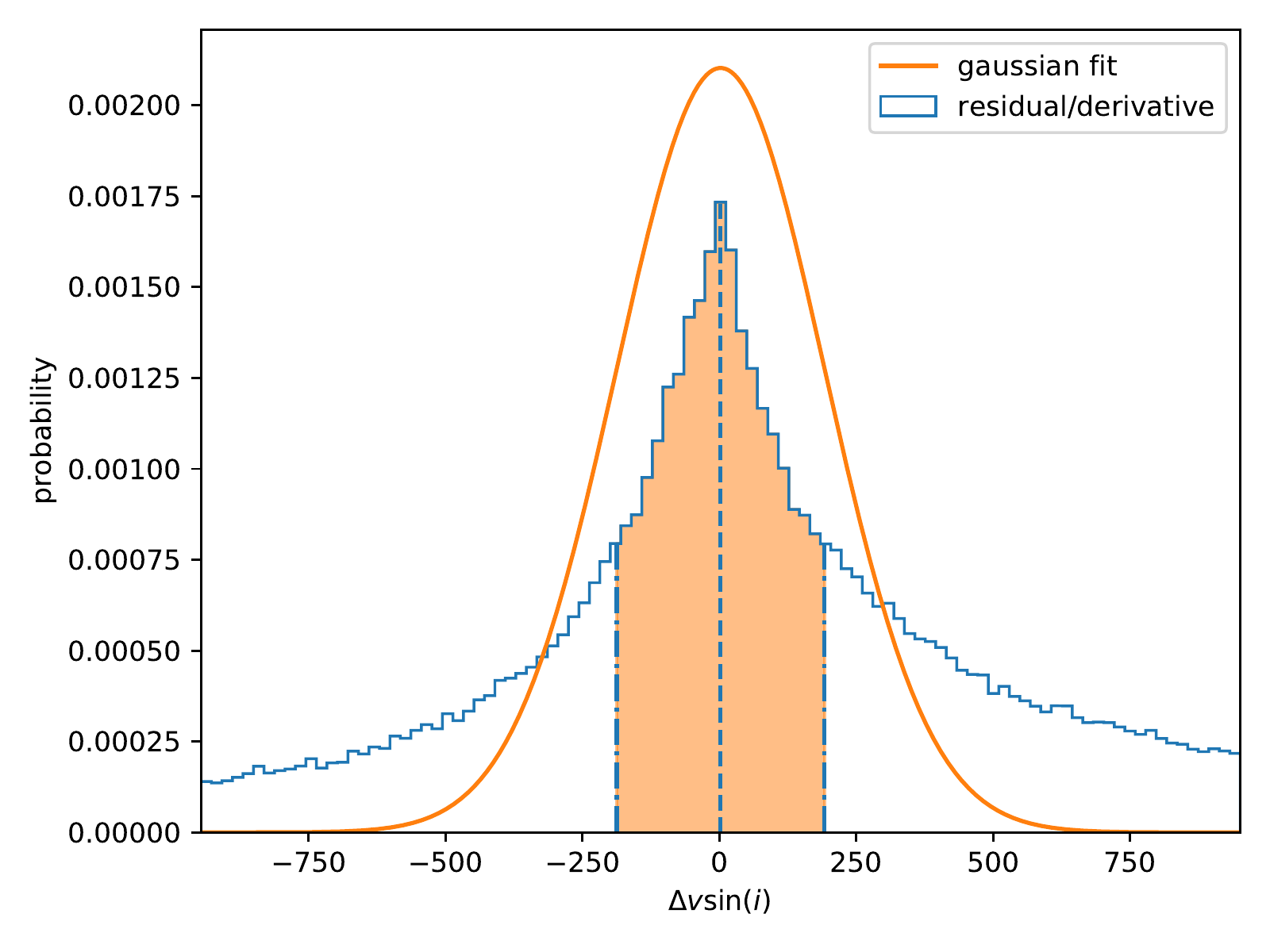}}
    \caption{Continuation of \autoref{fig:cum_prob_logg_app}}
    \label{fig:cum_prob_logg_app_2}
\end{figure*}

\clearpage

\section{Line List References}
\label{sec:linelist_references}

\begin{table*}[ht!]
    \begin{tabularx}{0.45\textwidth}{lX}
        \toprule
        Species & Reference \\
        \midrule
        H 1 & \citealt{CDROM18} \\
        Li 1 & \citealt{YD} \\
        C 1 & \citealt{NIST10,BPM} \\
        O 1 & \citealt{WSG} \\
        Na 1 & \citealt{NIST10,BPM,WSG,KP} \\
        Mg 1 & \citealt{LV,NIST10,AZS,LZ,BPM,KP,GUES} \\
        Mg 2 & \citealt{KP} \\
        Al 1 & \citealt{WSM,K75} \\
        Si 1 & \citealt{GARZ,K07} \\
        Si 2 & \citealt{K14} \\
        S 1 & \citealt{K04,WSM,BQZ,LWa} \\
        K 1 & \citealt{WSM} \\
        Ca 1 & \citealt{SR,K07,S,BPM,SN,DIKH} \\
        Ca 2 & \citealt{K10,TB} \\
        Sc 1 & \citealt{K09,BPM,LD,LHSNWC} \\
        Sc 2 & \citealt{LHSNWC,K09,LD} \\
        Ti 1 & \citealt{K16,BPM,KL} \\
        Ti 2 & \citealt{K16,MFW} \\
        V 1 & \citealt{MFW,K09,BPM,LWDFSC} \\
        V 2 & \citealt{K10,WLDSC,BGF} \\
        Cr 1 & \citealt{SLS,K16,BPM,WLHK,MFW} \\
        Cr 2 & \citealt{SLd,NLLN,PGBH,K16,GNEL,MFW} \\
        Mn 1 & \citealt{K07,DLSSC,BXPNL,BPM,MFW} \\
        Mn 2 & \citealt{K09} \\
        Fe 1 & \citealt{BWL,BPM,K14,FMW} \\
        Fe 2 & \citealt{K13,FMW} \\
        Co 1 & \citealt{BPM,LWG,K08,FMW} \\
        Co 2 & \citealt{K06} \\
        Ni 1 & \citealt{WLSCow,FMW,BPM,WLa,K08} \\
        Cu 1 & \citealt{K12} \\
         & \\
        \bottomrule
    \end{tabularx}
    \quad
    \begin{tabularx}{0.45\textwidth}{lX}
        \toprule
        Species & Reference \\
        \midrule
        Zn 1 & \citealt{LMW,Wa} \\
        Ge 1 & \citealt{LCG} \\
        Sr 1 & \citealt{CB,GC,PRT,VGH} \\
        Y 1 & \citealt{K06} \\
        Y 2 & \citealt{BBEHL,K11} \\
        Zr 1 & \citealt{CB,BGHL,MULT} \\
        Zr 2 & \citealt{MULT,LNAJ,CCout,BGHL,CC} \\
        Nb 1 & \citealt{DLa} \\
        Mo 1 & \citealt{WBb} \\
        Ru 1 & \citealt{WSL} \\
        Pd 1 & \citealt{CB} \\
        Cd 1 & \citealt{AS} \\
        In 1 & \citealt{PSa} \\
        Ba 1 & \citealt{CB,MW} \\
        Ba 2 & \citealt{MW,BPM} \\
        La 2 & \citealt{LBS,CB,ZZZ} \\
        Ce 1 & \citealt{CB,CBcor} \\
        Ce 2 & \citealt{PQWB,LSCI} \\
        Pr 1 & \citealt{MC} \\
        Pr 2 & \citealt{BLQS,ILW,MC} \\
        Nd 1 & \citealt{MC} \\
        Nd 2 & \citealt{HLSC,XSCL,MC} \\
        Sm 1 & \citealt{MC} \\
        Sm 2 & \citealt{MC} \\
        Eu 1 & \citealt{DHWL} \\
        Eu 2 & \citealt{LWHS} \\
        Gd 1 & \citealt{MC} \\
        Gd 2 & \citealt{DLSC,MC} \\
        Tb 2 & \citealt{LWCS} \\
        Dy 1 & \citealt{WLN} \\
        Dy 2 & \citealt{WLN,MC} \\
        Er 1 & \citealt{MC} \\
        Er 2 & \citealt{LSCW,MC} \\
        Lu 1 & \citealt{FDLP} \\
        Lu 2 & \citealt{QPBM,DCWL} \\
        Hf 1 & \citealt{CBcor,DSLb} \\
        W 1 & \citealt{OK} \\
        Re 1 & \citealt{DSLc} \\
        Os 1 & \citealt{CBcor} \\
        Pt 1 & \citealt{DHL} \\
        Tl 1 & \citealt{CB} \\
        C2 1 & \citealt{BBSB} \\
        CH 1 & \citealt{JLIY} \\
        MgH 1 & \citealt{KMGH} \\
        TiO 1 & \citealt{PPN2012} \\
        \bottomrule
    \end{tabularx}
    \caption{References for the input parameters of the line list used, sorted by element}
     \label{tab:linelist_reference}
\end{table*}

\clearpage

\section{NLTE Grid References}
\label{sec:nlte_references}

\begin{table}[ht]
    \centering
    \begin{tabularx}{\columnwidth}{lX}
         \toprule
         Element & Reference \\
         \midrule
         H &  {\citealt{2018A&A...615A.139A,2020A&A...642A..62A}} \\
         Li & {\citealt{2020A&A...642A..62A,2021MNRAS.500.2159W}} \\
         C & {\citealt{2019A&A...624A.111A,2020A&A...642A..62A}} \\
         N & {\citealt{2011A&A...528A.103L,2020A&A...642A..62A}} \\
         O & {\citealt{2018A&A...616A..89A,2020A&A...642A..62A}} \\
         Mg & {\citealt{2015A&A...579A..53O,2020A&A...642A..62A}} \\
         Al & {\citealt{2017A&A...607A..75N,2020A&A...642A..62A}} \\
         Si & {\citealt{2017MNRAS.464..264A,2020A&A...642A..62A}} \\
         K & {\citealt{2019A&A...627A.177R,2020A&A...642A..62A}} \\
         Ca & {\citealt{2019A&A...623A.103O,2020A&A...642A..62A}} \\
         Mn & {\citealt{2019A&A...631A..80B,2020A&A...642A..62A}} \\
         Fe & {\citealt{2016MNRAS.463.1518A}} \\
         Ba & {\citealt{2020A&A...634A..55G,2020A&A...642A..62A}} \\
         \bottomrule
    \end{tabularx}
    \caption{NLTE departure coefficient grid references}
    \label{tab:nlte_reference}
\end{table}

\end{document}